\documentclass[reprint,superscriptaddress,nofootinbib,amsmath,amssymb,aps]{revtex4-2}
\usepackage{graphicx}
\usepackage{dcolumn}
\usepackage{bm}
\usepackage{hyperref}
\hypersetup{colorlinks,citecolor=blue,linkcolor=blue,urlcolor=blue,}

\begin{document}

\title{A theoretical perspective on the almost dark galaxy Nube: exploring the fuzzy dark matter model}
\author{Yu-Ming Yang}
\email{yangyuming@ihep.ac.cn}
 \affiliation{%
 Key Laboratory of Particle Astrophysics, Institute of High Energy Physics, Chinese Academy of Sciences, Beijing 100049, China}
\affiliation{
 University of Chinese Academy of Sciences, Beijing 100049, China 
}%
\author{Xiao-Jun Bi}
\email{bixj@ihep.ac.cn}
\affiliation{%
 Key Laboratory of Particle Astrophysics, Institute of High Energy Physics, Chinese Academy of Sciences, Beijing 100049, China}
\affiliation{
 University of Chinese Academy of Sciences, Beijing 100049, China 
}%
\author{Peng-Fei Yin}
\email{yinpf@ihep.ac.cn}
\affiliation{%
 Key Laboratory of Particle Astrophysics, Institute of High Energy Physics, Chinese Academy of Sciences, Beijing 100049, China}

\begin{abstract}
In recent astronomical observations, an almost dark galaxy, designated as Nube, has unveiled an intriguing anomaly in its stellar distribution. Specifically, Nube exhibits an exceptionally low central brightness, with the 2D half-light radius of its stars far exceeding the typical values found in dwarf galaxies, and even surpassing those observed in ultra-diffuse galaxies (UDGs). This phenomenon is difficult to explain within the framework of cold dark matter (CDM). Meanwhile, due to its ultralight particle mass, fuzzy dark matter (FDM) exhibits a de Broglie wavelength on the order of kiloparsecs under the typical velocities of galaxies. The interference between different modes of the FDM wave gives rise to fluctuations in the gravitational field, which can lead to the dynamical heating of stars within galaxies, resulting in an expansion of their spatial distribution. In this paper, we aim to interpret the anomalous stellar distribution observed in Nube as a consequence of the dynamical heating effect induced by FDM. Our findings suggest that a FDM particle mass around $1-2\times 10^{-23}$ eV can effectively account for this anomaly. And we propose that the FDM dynamical heating effect provides a new insight into understanding the formation of field UDGs.

\end{abstract}


\maketitle

\section{Introduction\label{introduction}}
One of the most perplexing enigmas in modern physics is the elusive nature of dark matter. Since its inception \cite{zwicky6spectral,rubin1970rotation}, a multitude of theoretical models have been proposed, yet a robust theoretical framework must align with the observed phenomena in astrophysics. Cold dark matter (CDM) has shown consistency with numerous observational outcomes on a large scale, such as the large-scale structure of the universe and the cosmic microwave background. Nevertheless, recent advancements in the observation of small-scale structures have unveiled phenomena that challenge the predictions of CDM, such as the core-cusp problem \cite{flores1994observational,moore1994evidence,de2010core}, the missing satellites problem \cite{moore1999dark,klypin1999missing,zavala2009velocity}, the too-big-to-fail problem \cite{boylan2011too,boylan2012milky,tollerud2014m31,garrison2014too}, and others. Several alternative models to CDM have been proposed, such as warm dark matter \cite{bode2001halo,schneider2017hints,lovell2017addressing}, self-interacting dark matter \cite{spergel2000observational,tulin2018dark,adhikari2022astrophysical}, and fuzzy dark matter (FDM) \cite{hu2000fuzzy,peebles2000fluid,hui2017ultralight,hui2021wave,matos2023short,Khlopov_1999}. While these models exhibit behavior consistent with CDM on large scales, their distinct properties on small scales make them potentially capable of addressing the small-scale issues faced by CDM.

During the investigation of dark matter at the galactic scale, a significant challenge arises from the potential overlap of effects demonstrated by dark matter and baryonic matter. To address this issue, galaxies with minimal stellar density can be utilized, effectively discounting the effects of baryonic matter. The so-called almost dark galaxies \cite{janowiecki2015almost,leisman2017almost,brunker2019enigmatic}, which have only HI detections and are missed in the optical catalogues of wide field surveys such as the Sloan Digital Sky Survey (SDSS),  have an extremely low stellar surface mass density (a few $M_\odot$pc$^{-2}$). Recently, an almost dark galaxy, named Nube \cite{montes2024almost}, was identified by the IAC Stripe82 Legacy Project, and some of its properties were measured by the Gran Telescopio Canarias and the Green Bank Telescope. Comparing with other dwarf galaxies and ultra-diffuse galaxies (UDGs), Nube appears quite unique due to its extremely low central surface brightness and the large extension of its stellar distribution.  In the framework of CDM, the formation of Nube is a perplexing phenomenon, which will be discussed in the following section.

As an alternative to CDM, FDM is composed of ultralight ($m\sim 10^{-22}$ eV) bosons \footnote{It is worth to mention that such a light dark matter particle cannot be a fermion due to the Pauli exclusion principle. This principle sets a lower mass limit for fermion dark matter, known as the Tremaine-Gunn bound. \cite{Tremaine:1979we}.}. Consequently, the de Brogile wavelength of a particle with a typical galactic halo velocity dispersion ($\sigma\sim 100$ km s$^{-1}$) is approximately $\lambda_\sigma \sim 1$ kpc.  This leads to the fact that in galactic halos, the occupation number of FDM particles in a de Broglie volume is so high that these particles can be described by a classical field \cite{hui2021wave,Guth:2014hsa}, similar to the electromagnetic field \cite{article}.  In the non-relativistic limit \cite{Guth:2014hsa}, the system is governed by the  Schr$\ddot{\text{o}}$dinger-Poisson  equation. In the realm of FDM, it is observed that on scales significantly larger than the de Broglie wavelength, FDM behaves similarly to CDM. However, as the scale decreases to smaller sizes, particularly on the order of $\lambda_\sigma$,  the quantum effects cause FDM to diverge from the behavior of CDM. For instance, the FDM halo comprises a solitonic core \cite{Schive_2014_1,Schive_2014}, representing the ground-state solution of the  Schr$\ddot{\text{o}}$dinger-Poisson  equation \cite{Chavanis_2011}, alongside an envelope of excited states that resembles the NFW profile \cite{Navarro_1997}.  The most distinctive property of FDM is the wave interference between different states \cite{Li_2021}, resulting in fluctuations of order unity within the NFW envelope \cite{Schive_2014,Liu_2023}, soliton oscillations \cite{Veltmaat_2018}, and soliton random walk \cite{Schive_2020}. These properties contribute to fluctuations in the gravitational field, potentially leading to the dynamical heating of objects within the halo \cite{Bar_Or_2019,El_Zant_2019,chavanis2020landau}, resulting in phenomena such as the expansion of the stellar distribution \cite{Dutta_Chowdhury_2023} and the ejection of nuclear objects from the center of the halo \cite{Dutta_Chowdhury_2021}. Previous studies have utilized the dynamical heating effect to constrain the mass of FDM particles \cite{Marsh_2019,Church_2019,2022PhRvD.106f3517D,Chiang_2021}, although certain problematic issues remain.

In this study, we aim to explain the anomalous stellar distribution observed in Nube as a consequence of the dynamical heating effect induced by FDM. To achieve this, we utilize the quasiparticles perspective developed and the diffusion coefficients derived in \cite{Bar_Or_2019} to calculate the orbit radius expansion of each particle\footnote{Here, the term “particle" refers to the object utilized to represent the stars in our calculations.}. Additionally, we incorporate the heating from the soliton oscillations and random walk using the effective model proposed in \cite{Dutta_Chowdhury_2021}, which has been demonstrated to align well with simulations. Our approach assumes that   the distribution of stars at the start of simulation is described by a Plummer profile \cite{1911MNRAS..71..460P} with the radius parameter of 1 kpc.  We propose a sampling method to account for the randomness of the energy increase of each particle, and we have verified that it aligns well with the simulation results in \cite{Dutta_Chowdhury_2023}.  We use this model to evolve each particle through the age of Nube and then analyze to obtain the 2D stellar mass surface density distribution today. Our findings indicate that a FDM particle mass of approximately $1-2\times 10^{-23}$ eV can effectively explain the anomaly of Nube. While this mass range is consistent with some results in other studies, it is in tension with certain exclusion bounds, such as those inferred from the Lyman-$\alpha$ forest, subhalo mass function, and dynamical heating effect in dwarf galaxies. We intend to address the uncertainties in these studies.

Furthermore, we propose that the FDM dynamical heating effect may provide a new insight into understanding the formation of field UDGs, and qualitatively aligns with the existing observational outcomes. From this perspective, Nube could be considered a distinctive field UDG, having not been accreted into a host halo to become a cluster UDG, but instead persisting in the field for a long duration.

The paper is organized as follows. In Sec.~\ref{observation} we present the observations of Nube and address the anomaly of the stellar distribution in Nube compared to other dwarf galaxies and UDGs. In Sec.~\ref{FDM} we introduce a theoretical background of the FDM and describe the dynamical heating effect induced by wave interference. The analysis of Nube is performed in Sec.~\ref{nube}. Finally, we discuss our results and provide a conclusion in Sec.~\ref{discussions} and Sec.~\ref{conclusion}, respectively.

\section{Observations of Nube\label{observation}}
As an almost dark galaxy, Nube is not visible in SDSS images and was instead serendipitously discovered in the IAC Stripe82 Legacy Project \cite{montes2024almost}. Through an analysis of the 100 m Green Bank Telescope HI data, Nube was determined to be located at a distance of $107\pm5$ Mpc. Additionally, the surface stellar mass density profile was constructed using images from the 10.4 m Gran Telescopio Canarias HiPERCAM across five different bands. This density profile enabled the calculation of the stellar 2D half-light radius, which was found to be $R_e=6.9\pm 0.8$ kpc, and the total stellar mass, which was determined to be $M_\star=(3.9\pm 1.0)\times 10^8M_\odot$. The age of Nube was estimated to be $10.2^{+2.0}_{-2.5}$ Gyr. Furthermore, applying the equation from \cite{2018ApJ...855...28S}, the dynamical mass of Nube within $3R_e$ was calculated to be $(2.6\pm 1.7)\times 10^{10}M_\odot$. 

In comparison to other galaxies with similar masses, Nube exhibits a relatively flat stellar mass density profile and lower central surface brightness. Furthermore, its 2D half-light radius is even larger than the typical values of UDGs ($1.5\sim 5$ kpc). While the UDGs can be produced in the CDM simulations through bayonic
feedback \cite{di2017nihao,chan2018origin}, tidal stripping and heating \cite{carleton2019formation,jiang2019formation,sales2020formation}, and high spin effect \cite{amorisco2016ultradiffuse}, the observed properties of Nube cannot be produced in these scenarios \cite{montes2024almost}. In addition, there are observational signs indicating that Nube has not been subjected to significant tidal forces, indicating that it is not a tidal dwarf galaxy, nor is it formed by its interaction with the environment \cite{montes2024almost}. Therefore, within the framework of CDM, the formation of Nube presents a perplexing phenomenon.

\section{Theoretical description of FDM\label{FDM}}
\subsection{Halo density profile}
The behavior of FDM can be effectively described by a classical, non-relativistic field, and the impact of the Hubble expansion can be disregarded on the galactic scale. Consequently, the equation of motion governing FDM is represented by the  Schr$\ddot{\text{o}}$dinger-Poisson  system, which is expressed by the following equations in natural units ($\hbar=c=1$)
\begin{equation}
\begin{aligned}
    &i\partial_t \psi=-\frac{\nabla^2}{2m}\psi+m\Phi\psi,\\
    &\nabla^2\Phi=4\pi G\rho, \quad \rho=m\left|\psi\right|^2.
\end{aligned}
\end{equation}
The ground-state solution of this system manifests as a spherically symmetric soliton \cite{Chavanis_2011}, which was also observed in simulations \cite{Schive_2014_1,Schive_2014}. The density profile of the soliton can be accurately represented by \cite{Schive_2014_1}
\begin{equation}
    \rho_\text{sol}(r)=\frac{\rho_c}{\left[1+0.091(r/r_c)^2\right]^8},
\end{equation}
where $\rho_\text{c}$ represents the soliton core density, and $r_\text{c}$ is the core radius, defined such that $\rho_\text{sol}(r_\text{c})=\rho_\text{c}/2$. The density of the soliton core and the radius of the core are related by the scaling symmetry of the  Schr$\ddot{\text{o}}$dinger-Poisson   equation \cite{Guzman_2006,Mocz_2017}
\begin{equation}
    \rho_c=1.95\times 10^7M_\odot \text{kpc}^{-3}\left(\frac{m}{10^{-22}\text{eV}}\right)^{-2}\left(\frac{r_c}{\text{kpc}}\right)^{-4}.
    \label{rho_c_r_c}
\end{equation}

A realistic FDM halo is characterized by a solitonic core surrounded by a NFW-like envelope made up of excited states. The gravitational equilibrium of the FDM halo is primarily maintained by quantum pressure in the soliton core, while in the NFW region, half of the gravitational force is balanced by random motion and the other half by quantum pressure \cite{Mocz_2017,Dutta_Chowdhury_2021}. The transition radius from the soliton to the NFW region is proportional to the core radius $r_\text{c}$, with the proportional coefficient varying in different studies, such as $2.7$ in \cite{Dutta_Chowdhury_2021}, $3.3$ in \cite{Chiang_2021}, $3.5$ in \cite{Mocz_2017}. In the subsequent analysis, we define the transition radius as $kr_\text{c}$, where $k$ is a parameter that ranges from 2.7 to 3.5. Consequently, the time-averaged density profile of a FDM halo is given by
\begin{equation}
    \rho_\text{FDM}(r)=\left\{\begin{array}{ll}
        \frac{\rho_c}{\left[1+0.091 (r/r_c)^2\right]^8} &,\; r<kr_c\\
        \frac{\rho_s}{\frac{r}{r_s}\left(1+\frac{r}{r_s}\right)^2} &,\; r\geq kr_c.
    \end{array}\right.
\label{rho_FDM}
\end{equation}

\subsection{Dynamical heating}
The interference between excited states within the NFW envelope induces density fluctuations of approximately unity order, giving rise to various structures such as vortex lines (regions of zero density) and granules with sizes on the order of the de Broglie wavelength $\lambda_\sigma$ \cite{Hui_2021,Liu_2023,Mocz_2017}. These structures evolve over a timescale of $\lambda_\sigma/\sigma$, resulting in persistent fluctuations in the gravitational field. Additionally, in the soliton region, the interference between the ground-state and excited states leads to soliton oscillations and random walk \cite{Li_2021}, which can also cause fluctuations in the gravitational field. The inhomogeneous gravitational field accelerates particles within the halo according to Newton's law, with the velocity and energy changes being described by the velocity diffusion coefficient and the energy diffusion coefficient, respectively.

In the NFW envelope, the heating effect can be described according to \cite{Bar_Or_2019}, where the authors demonstrated that the diffusion coefficients in FDM bear similarities to the classical two-body relaxation case. Specifically, FDM behaves like quasiparticles with an effective mass $m_\text{eff}$, which depends on the FDM particle's de Broglie wavelength, as well as its density and velocity distribution. For the Maxwellian velocity distribution, the effective mass is given by
\begin{equation}
    m_\text{eff}=\frac{\pi^{3/2} \rho_\text{FDM}}{m^3\sigma^3}=\rho_\text{FDM}\left(\frac{\lambda_\sigma}{2\sqrt{\pi}}\right)^3,
\end{equation}
where $\sigma$ is equal to the velocity dispersion derived from solving the Jeans equation \cite{Dutta_Chowdhury_2021}, and $\lambda_\sigma = 1/m\sigma$ is the typical de Broglie wavelength.
For a zero-mass test particle in NFW envelope, its first and second order velocity diffusion coefficients\footnote{The diffusion coefficients are defined as $D[\Delta v_i]\equiv \langle\Delta v_i\rangle/\Delta t$,$D[\Delta v_i\Delta v_j]\equiv \langle\Delta v_i\Delta v_j\rangle/\Delta t$, where $\Delta v_i$ represents the change in the velocity component $i$ over a time interval $\Delta t$, and $\langle\cdot \rangle$ denotes the ensemble average.} are given by
\begin{equation}
D[\Delta v_\parallel]=-\frac{4\pi G^2 \rho_\text{FDM}m_\text{eff}\ln\Lambda_\text{FDM}}{\sigma^2_\text{eff}}\mathbb{G}(X_\text{eff}),
\label{D_1}
\end{equation}
\begin{equation}
    D[(\Delta v_\parallel)^2]=\frac{4\sqrt{2}\pi G^2 \rho_\text{FDM}m_\text{eff}\ln\Lambda_\text{FDM}}{\sigma_\text{eff}}\frac{\mathbb{G}(X_\text{eff})}{X_\text{eff}},
\label{D_2}
\end{equation}
\begin{equation}
    D[(\Delta v_\perp)^2]=\frac{4\sqrt{2}\pi G^2 \rho_\text{FDM}m_\text{eff}\ln\Lambda_\text{FDM}}{\sigma_\text{eff}}\frac{\text{erf}(X_\text{eff})-\mathbb{G}(X_\text{eff})}{X_\text{eff}}.
\label{D_3}
\end{equation}
Here, the $\Delta v_\parallel$ and $\Delta v_\perp$ represent the velocity changes parallel and perpendicular to the test particle's velocity $\mathbf{v}$, respectively.  Additionally, $\sigma_\text{eff}=\sigma/\sqrt{2}$ is the effective FDM particle velocity dispersion, and $X_\text{eff}\equiv v/\sqrt{2}\sigma_\text{eff}$, where $v$ is the test particle's velocity. The Coulomb logarithm, denoted as $\ln \Lambda_\text{FDM}$, is  given by
\begin{equation}
    \ln\Lambda_\text{FDM}=\ln\left(\frac{4\pi r}{\lambda_\sigma}\right),
\end{equation}
where $r$ is the distance to the halo center. Furthermore, $\mathbb{G}$ is a function defined as
\begin{equation}
    \mathbb{G}(X)=\frac{1}{2X^2}\left[\text{erf}(X)-\frac{2X}{\sqrt{\pi}}e^{-X^2}\right].
\end{equation}

Using these velocity coefficients, we can derive the expression for the energy diffusion coefficient, which describes the ensemble-averaged change rate of the energy per unit mass of the test particle
\begin{equation}
    \begin{aligned}
        D[\Delta E]&\equiv \frac{\langle \Delta E/M\rangle}{\Delta t}\\
        &=\frac{\langle (\mathbf{v}+\Delta \mathbf{v})^2-\mathbf{v}^2\rangle }{2\Delta t}\\
        &=vD[\Delta v_\parallel]+\frac{1}{2}D[(\Delta v_\parallel)^2]+\frac{1}{2}D[(\Delta v_\perp)^2]\\
        &=\frac{4\sqrt{2\pi}G^2 \rho_\text{FDM}m_\text{eff}\ln\Lambda_\text{FDM}}{\sigma_\text{eff}}e^{-X_\text{eff}^2}.
    \end{aligned}
    \label{D_Delta_E}
\end{equation}
Additionally, we can define $D[(\Delta E)^2]$ as 
\begin{equation}
\begin{aligned}
    D\left[(\Delta E)^2\right]&\equiv \frac{\langle(\Delta E/M)^2\rangle}{\Delta t}\\
    &=\frac{\langle\left(v \Delta v_\parallel+\frac{1}{2}(\Delta v_\parallel)^2+\frac{1}{2}(\Delta v_\perp)^2\right)^2\rangle}{\Delta t}\\
    &=v^2D[(\Delta v_\parallel)^2]\\
    &=8\sqrt{2}\pi G^2\rho_\text{FDM}m_\text{eff}\sigma_\text{eff}\ln\Lambda_\text{FDM} X_\text{eff}\mathbb{G}(X_\text{eff}).
\end{aligned}
\end{equation}

It is important to emphasize that the derivation of Eq.\,\ref{D_1}, Eq.\,\ref{D_2}, and Eq.\,\ref{D_3} is based on the assumption of an infinite, homogeneous system of FDM particles. However, in this study, we utilize a local approximation to apply them in a FDM halo, which is only valid for larger radii with $\Lambda_\text{FDM} \gtrsim 1$ \cite{Chiang_2022}. Moreover, at smaller radii, the dynamical heating induced by soliton oscillations and soliton random walk predominates over the heating described above. In \cite{Dutta_Chowdhury_2021}, the authors proposed  an effective model to incorporate the contributions from the soliton. Specifically, they replace the values of the diffusion coefficients at all the locations within $2.3r_c$ with the value at $2.3r_c$, and found that this effective model can fit the simulation results quite well in the case of zero-mass test particles. We adopt this model in our subsequent analysis and have verified that all our calculations are within the regime of $\Lambda_\text{FDM} >1$.

\section{Interpretation of the anomaly in Nube with FDM\label{nube}}
\subsection{FDM halo density profile and initial stellar distribution}
The time-averaged FDM halo density profile is described by Eq.\,\ref{rho_FDM}, which involves five parameters: $\rho_c,r_c,\rho_s,r_s$, and $k$. However, the parameters $\rho_c$ and $r_c$ are related by Eq.\,\ref{rho_c_r_c}, and the continuity condition at $kr_c$ also relates $\rho_s$ and $\rho_c$ 
\begin{equation}
    \frac{\rho_c}{(1+0.091k^2)^8}=\frac{\rho_s}{\frac{kr_c}{r_s}\left(1+\frac{kr_c}{r_s}\right)^2}.
\end{equation}
Additionally, there is a constraint from the dynamical mass within $3R_e$ of Nube\footnote{  In this study, the parameters of Nube as outlined in \cite{montes2024almost}, including the dynamical mass within $3R_e$, age, 2D half-light radius, and total stellar mass, are assumed to be at their optimal values unless otherwise specified. }
\begin{equation}
    M_\text{FDM}(3R_e)=\int_0^{3R_e} 4\pi r^2\rho_\text{FDM}(r)dr=2.6\times 10^{10} M_\odot,
\end{equation}
where the contribution from stellar and gas components are omitted due to their total mass being much smaller than that of dark matter. Consequently, with the above three constraints, there are only two free parameters, which we choose to be $k$ and $r_s$, with the assumption that $k\in [2.7,3.5]$, and $r_s\in [5,20]$ kpc, with default values of $k=3,r_s=10$ kpc.  The impact of these parameters on the results will be discussed in the following sections.

The initial stellar mass density distribution is modeled using a Plummer profile, which is known to provide a good fit to the distribution of stars in dwarf galaxies \cite{Geringer_Sameth_2015,Koushiappas_2017}. The profile is described by the equation
\begin{equation}
    \rho_\star(r)=\frac{3M_\star}{4\pi a_i^3}\left(1+\frac{r^2}{a_i^2}\right)^{-\frac{5}{2}},
    \label{Plum}
\end{equation}
where $M_\star=3.9\times 10^8 M_\odot$ is the total stellar mass, and the Plummer radius $a_i$ is equal to the initial 2D half-light radius,  typically falling within the range of $\lesssim 1.5$ kpc for classical dwarf galaxies. We assume a range of $a_i\in[0.5,1.5]$ kpc, and find that varying the value of $a_i$ within this range has almost no impact on the final stellar density profile. Therefore, we fix $a_i$ to be $1$ kpc in the following analysis.

\subsection{Stellar mass density profile evolution}
Considering that a particle's orbit is typically non-circular, we define the “mean" radius $r$ of a particle's orbit  as the radius of a circular orbit with equivalent gravitational potential energy. This radius can be considered to represent the average location of the particle, and we simply state that it is located at $r$. Under the assumption of virialization, the total energy per unit mass this particle can be expressed as
\begin{equation}
    \frac{E}{M}=\frac{1}{2}\frac{U}{M}=-\frac{1}{2}\int_r^\infty \frac{GM_\text{FDM}(r^\prime)}{{r^\prime}^2}dr^\prime,
    \label{virial}
\end{equation}
where the contributions from stellar and gas components have been omitted. The dynamical heating from the FDM results in an energy change $\Delta E$ to this particle within a time interval $\Delta t$, leading to a change in the radius of its orbit as described by 
\begin{equation}
    \Delta r=\frac{2 r^2}{GM_\text{FDM}(r)}\frac{\Delta E}{M}.
    \label{Delta_r}
\end{equation}
Based on Eq. \ref{D_Delta_E}, the ensemble-averaged value of $\Delta E/M$ is given by
\begin{equation}
    \langle\Delta E/M\rangle=D[\Delta E]\Delta t.
    \label{averageEM}
\end{equation}
Furthermore, the variance of $\Delta E/M$ can be derived as
\begin{equation}
    \begin{aligned}
    \sigma^2(\Delta E/M)&=\langle\left(\Delta E/M-\langle\Delta E/M\rangle\right)^2\rangle\\
    &=\langle(\Delta E/M)^2\rangle-\langle\Delta E/M\rangle^2\\
    &=D\left[(\Delta E)^2\right]\Delta t-D[\Delta E]^2\Delta t^2
    \end{aligned}.
\end{equation}

\begin{figure*}
  \includegraphics[width=0.49\textwidth]{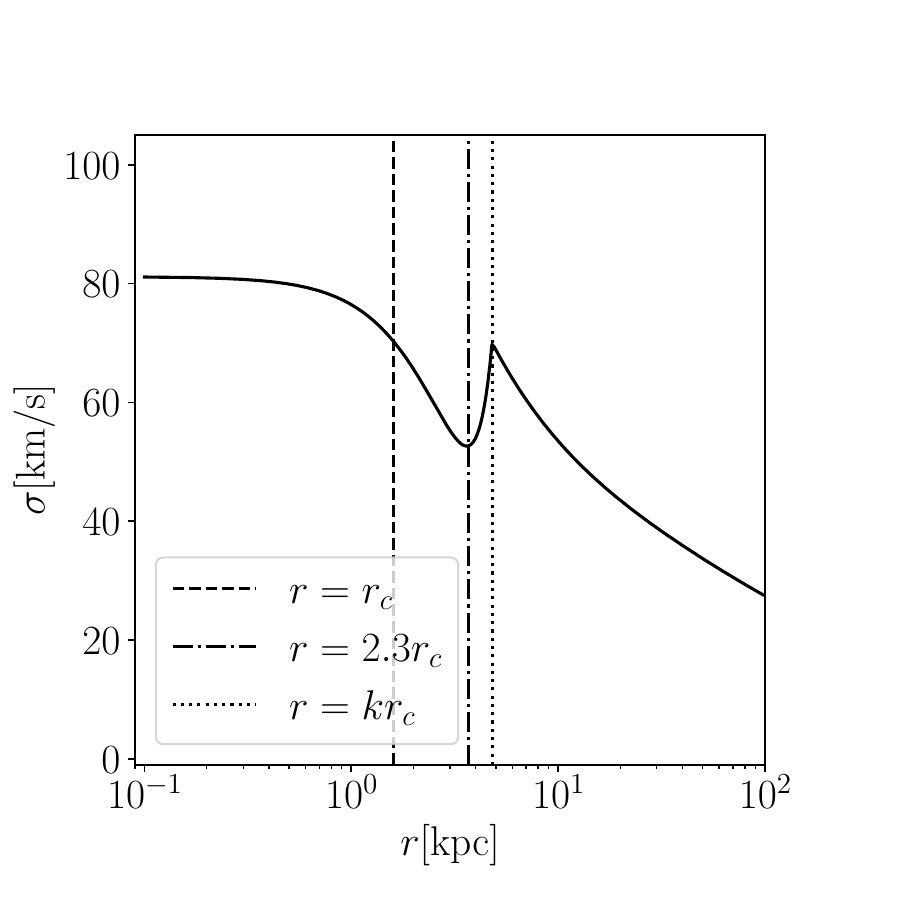}
  \hfill
  \includegraphics[width=0.49\textwidth]{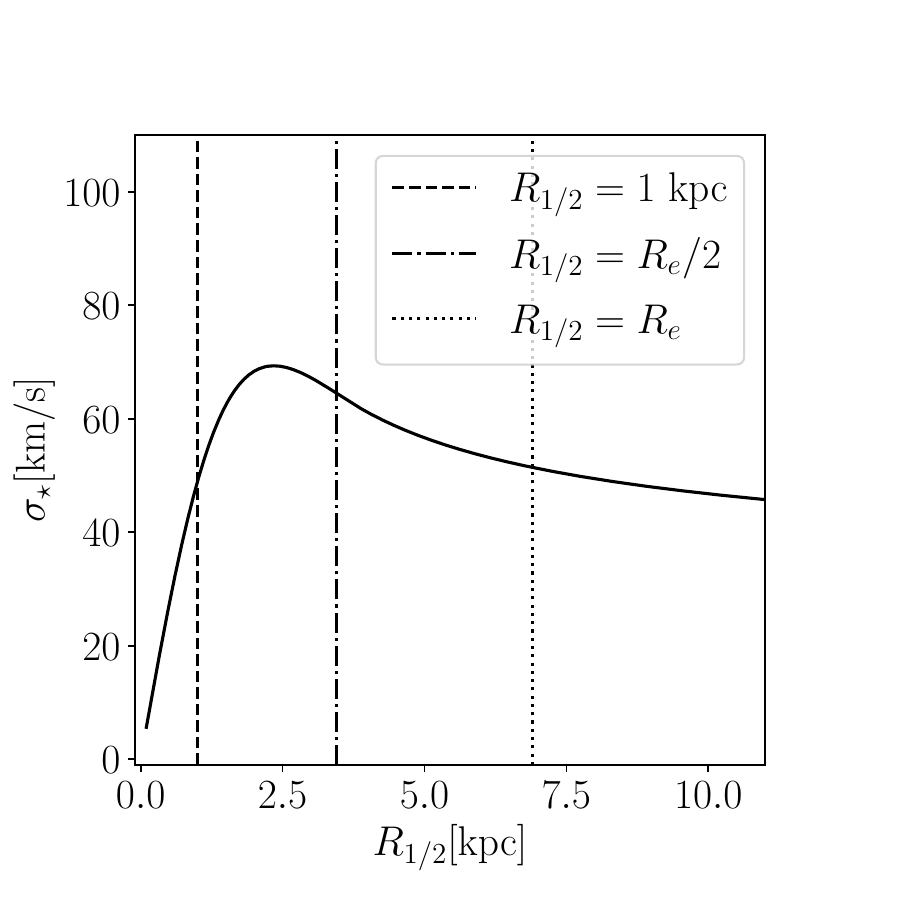}
  \caption{Left panel: the solution of the isotropic Jeans equation. The three vertical lines indicate the positions of radius of radius $r=r_c$, $r=2.3r_c$, and $r=kr_c$, respectively. Right panel: the solution of $\sigma_\star$ as a function of $R_{1/2}$. The three vertical lines denoting $R_{1/2}=1$ kpc, $R_{1/2}=R_e/2$, and $R_{1/2}=R_e$, respectively.}
  \label{figure_1}
\end{figure*}
The values of $D[\Delta E]$ and $D[(\Delta E)^2]$ are dependent on the FDM 1D velocity dispersion $\sigma$ and the stellar velocity $v$. As stated in \cite{Dutta_Chowdhury_2021}, the effective FDM velocity dispersion $\sigma_\text{eff}$ outside the soliton  is connected to $\sigma$ through $\sigma_\text{eff}=\sigma/\sqrt{2}$, where $\sigma$ is derived by solving the Jeans equation \footnote{The genuine 1D halo velocity dispersion including quantum effects drived from simulation is related to $\sigma$ by $\sigma_h=\sigma/\sqrt{2}$ outside the soliton. See Ref.\,  \cite{Dutta_Chowdhury_2021} for a detailed discussion.}
\begin{equation}
    -\frac{1}{\rho_\text{FDM}}\frac{d\left(\rho_\text{FDM}\sigma^2\right)}{dr}-\frac{2\beta\sigma^2}{r}=\frac{d\Phi_\text{FDM}(r)}{dr}=\frac{GM_\text{FDM}(r)}{r^2}.
\end{equation}
Assuming isotropy ($\beta=0$), the solution with $m=1.0\times 10^{-23}$ eV and default values of $k,r_s$ is illustrated in the left panel of Fig.\,\ref{figure_1}. In subsequent calculations, we neglect the radius dependence of $\sigma$ and set it to be the value at $kr_c$.  The region we are most interested in is about $2.3r_c\leq r\lesssim 10 \text{kpc}$, where the deviation of $\sigma$ from the default value mentioned above is at most about $ -24\%$. We will delve into the impact of this uncertainty on the output in subsequent discussions.

The stellar velocity can be related to its 1D velocity dispersion by $v\sim \sqrt{3}\sigma_\star$. We approximate the value of $\sigma_\star$ using the solution of the equation \cite{Wolf_2010,Simon_2019}
\begin{equation}
    M_\text{FDM}\left(\frac{4}{3}R_{1/2}\right)=9.3\times 10^5\left(\frac{\sigma_\star}{\text{km/s}}\right)^2\left(\frac{R_{1/2}}{\text{kpc}}\right)M_\odot.
\end{equation}
The solution of $\sigma_\star$ as a function of $R_{1/2}$ is shown in the right panel of Fig.\,\ref{figure_1} with the same parameters chosen as the left panel. In our analysis, the $R_{1/2}$ is not a constant, but evolves with time, so we select $\sigma_\star$ to be its value at $R_{1/2}=R_e/2$.  Within the range of $R_{1/2}$ the star distribution experienced ($1\;\text{kpc}\lesssim R_{1/2}\lesssim 6.9 \;\text{kpc}$), the deviation of $\sigma_\star$ from the default value mentioned above is at most about $ -22\%$. Similar to the uncertainty associated with $\sigma$, we will also address the impact of the uncertainty of $\sigma_\star$.  

At the end of each time bin, we sample an energy change $\Delta E/M$ from a Gaussian distribution with mean value and standard deviation of $\langle\Delta E/M\rangle$ and $\sigma(\Delta E/M)$, respectively. Subsequently, we update the radius of the particle's orbit according to Eq.\,\ref{Delta_r}. However, a cut-off is implemented at a radius of $0.1$ kpc; if the updated radius is smaller than $0.1$ kpc, it is manually adjusted to be $0.1$ kpc.

The total number of particles and the time step are set to be $N=10^6$ and $\Delta t=1$ Myr, respectively. It has been verified that the results remain consistent even when using a larger particle number and a smaller time step.Each particle is evolved for a duration of $10.2$ Gyr based on the aforementioned procedure, and the final stellar mass distribution is obtained by statistically analyzing the locations of all particles.

\begin{figure*}
    \includegraphics[width=0.49\textwidth]{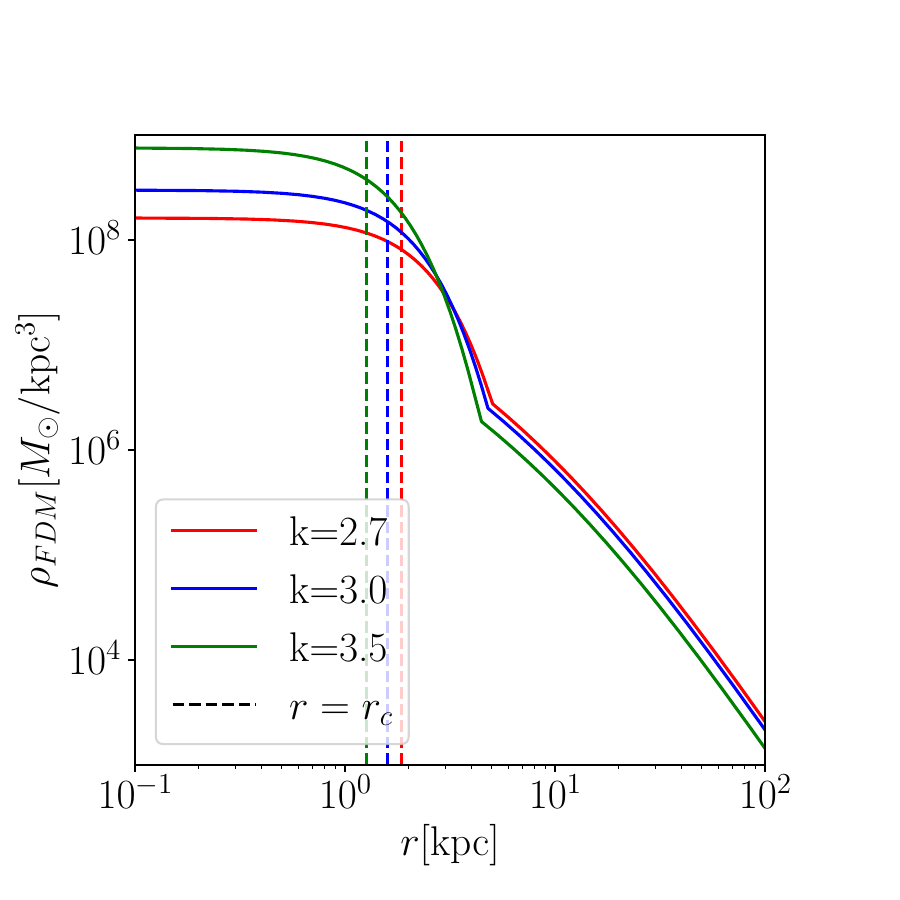}
  \hfill
    \includegraphics[width=0.49\textwidth]{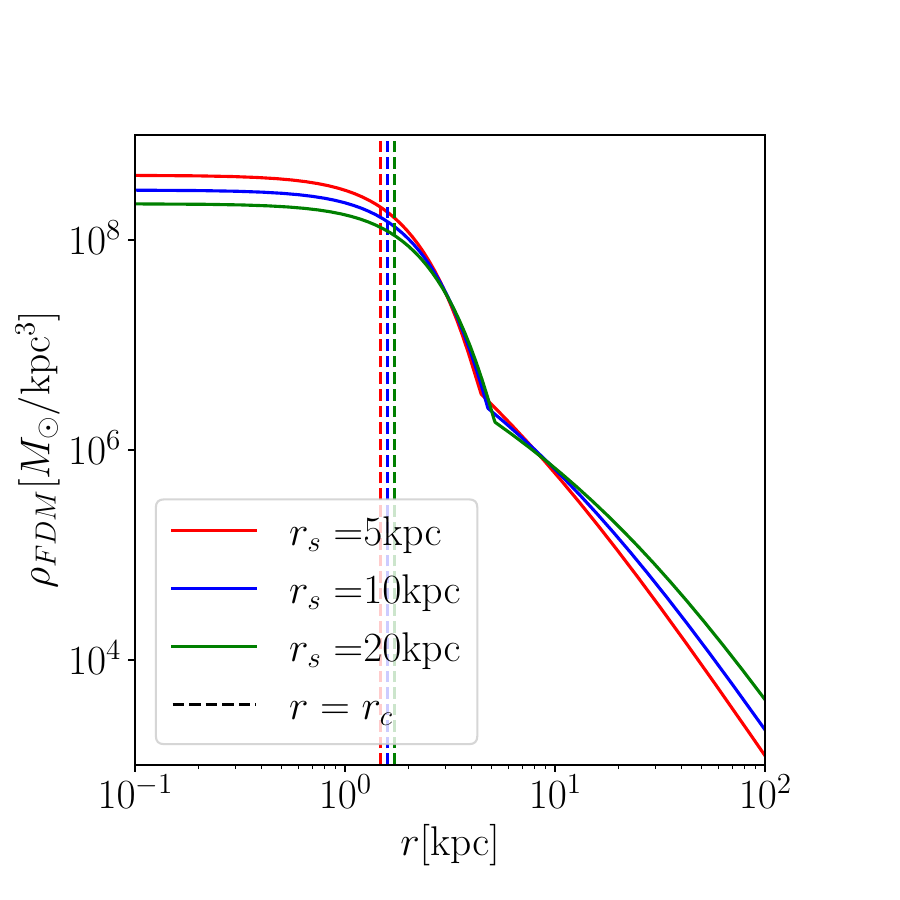}
    \includegraphics[width=0.49\textwidth]{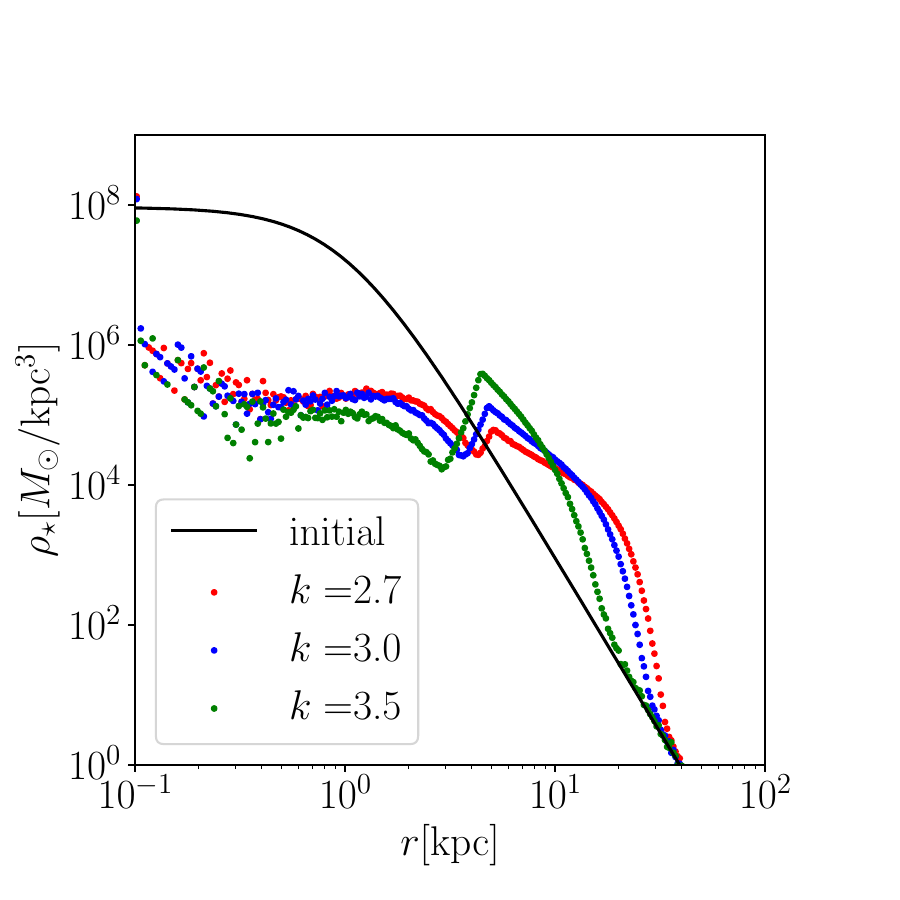}
  \hfill
    \includegraphics[width=0.49\textwidth]{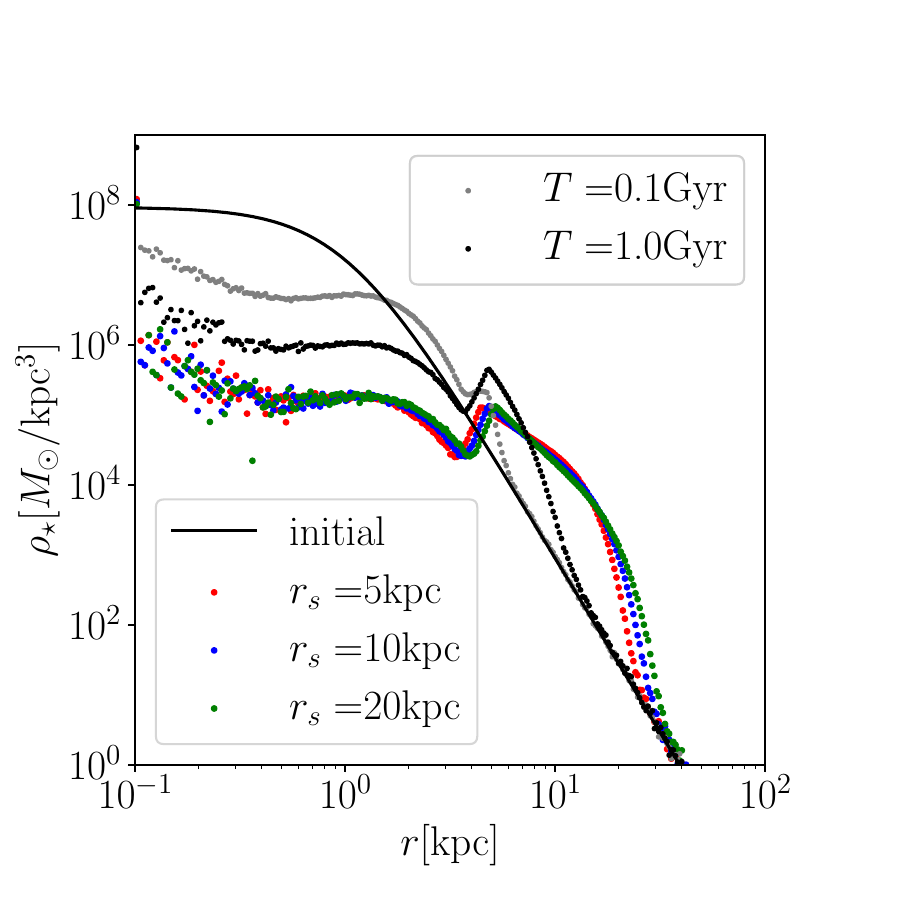}
  \caption{Upper panels: the impact of $k$ and $r_s$ on the FDM density profile, with the vertical dashed lines representing the position of $r=r_c$. Lower panels: the impact of $k$ and $r_s$ on the final 3D stellar mass density profile, with the black solid line representing the initial profile, and the  red, blue, and green  dots indicating the final density in each radius bin. The default parameter values are $k=3$ and $r_s=10$ kpc unless otherwise specified, and the FDM particle mass $m=1.0\times 10^{-23}$ eV.  For the default values of $k$ and $r_s$, two snapshots at $T=0.1$ Gyr and $T=1.0$ Gyr are depicted using grey and black dots, respectively, in the lower right panel.}
  \label{figure_2}
\end{figure*}

The impact of parameters $k$ and $r_s$ is depicted in Fig.\,\ref{figure_2}, with the default values of $k=3$, $r_s=10$ kpc if not otherwise specified, and the FDM particle mass $m=1.0\times 10^{-23}$ eV. The upper panels display the FDM density profile, with the vertical dashed lines representing the position of $r=r_c$. It is evident that the impact of $k$ is significant, while the impact of $r_s$ is minimal. In the lower panels, the stellar mass density profile is displayed, with the black solid line representing the initial profile, and the red, blue, and green dots indicating the final density in each radius bin. In the region of $r\lesssim 0.3$ kpc, the number of particles in each bin is only one or zero, so the representation of the dots in this region is not realistic. In the subsequent analysis, we consider the density in this region to be constant, equivalent to the value at $r=0.3$ kpc.  As $k$ and $r_s$ can only influence the final stellar density through their impact on the FDM density, the impact of $k$ and $r_s$ on the stellar density is also significant and minimal, respectively, mirroring their impact on the FDM density. Consequently, we only discuss the impact of $k$ on the results and omit the impact of $r_s$ in the following section. 

 In the lower right panel of Fig.\,\ref{figure_2}, two snapshots at $T=0.1$ Gyr and $T=1.0$ Gyr are shown using grey and black dots, respectively, with the default values of $k$ and $r_s$. The temporal evolution of the stellar density exhibits a notable feature where the heating rate diminishes over time. This trend is consistent with the simulation outcomes reported in \cite{Dutta_Chowdhury_2023}. Additionally, the final stellar density distribution exhibits a distinct trough followed by a peak, as depicted in the lower panels of Fig.\,\ref{figure_2}. These structures stem from the effective model proposed in \cite{Dutta_Chowdhury_2021}. Further discussions on them are provided in Appendix \ref{App}.

\subsection{Fitting to the observational data} 
\begin{figure*}
    \includegraphics[width=0.49\textwidth]{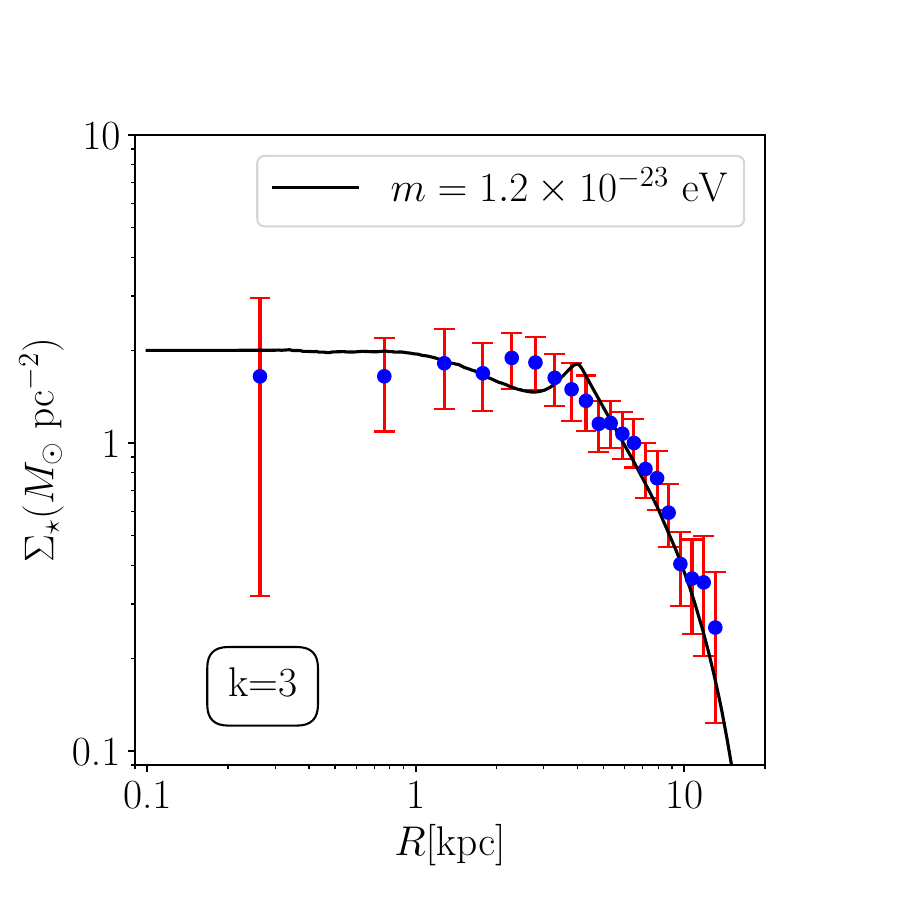}
  \hfill
    \includegraphics[width=0.49\textwidth]{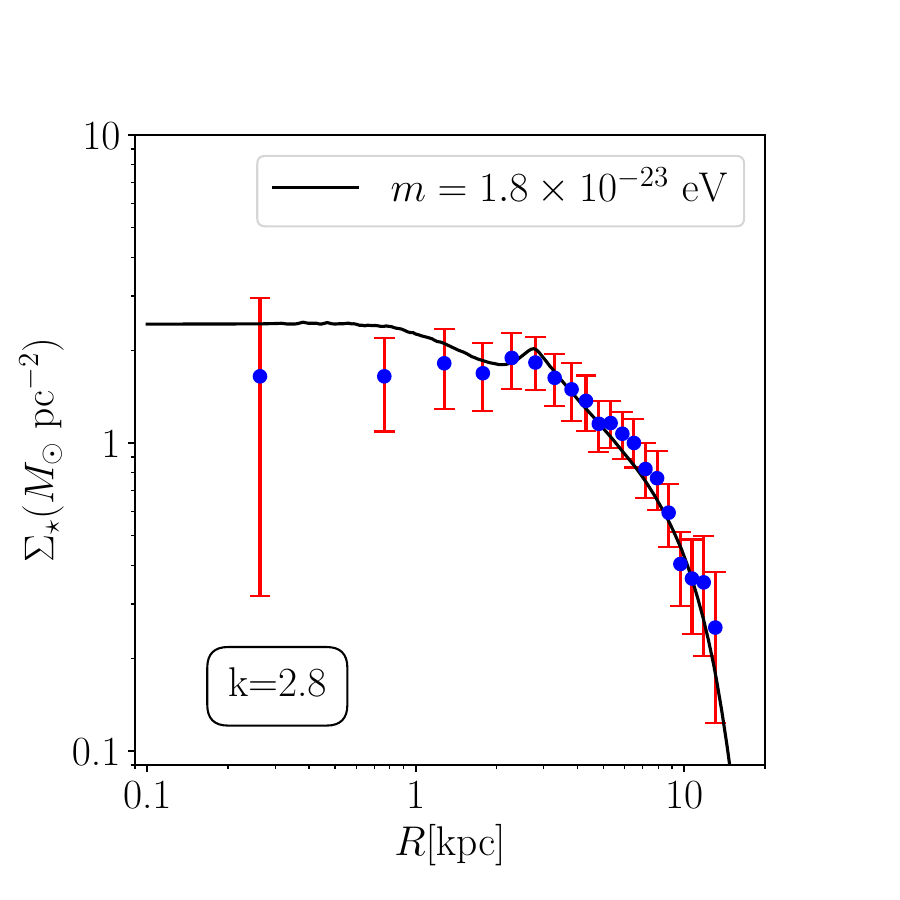}
  \caption{Left panel: the best fit profile with the default value of $k=3$, resulting in a corresponding FDM particle mass of $m=1.2\times 10^{-23}$ eV, and $\chi^2\simeq 16.8$. Right panel: the corresponding profile of the minimum value of $\chi^2$ obtained for $k = 2.8 $, with the FDM particle mass $m=1.8\times 10^{-23}$ eV, and $\chi^2\simeq 12.5$. }
  \label{figure_3}
\end{figure*}

Initially, we maintain $k$ and $r_s$ at their default values and adjust the FDM particle mass $m$ to generate a series of different final 3D density profiles as those depicted in the lower panels of Fig.\,\ref{figure_2}. These profiles were then converted to 2D surface density profiles, and the optimal value of $m$ that best fit the data was determined. The optimal mass was found to be $m=1.2\times 10^{-23}$ eV with $\chi^2\simeq 16.8$, and the correspond profile comparison with the observational data is shown in the left panel of Fig.\,\ref{figure_3}. 

 If we switch to using  the FDM velocity dispersion $\sigma$ at $r=2.3r_c$,  as shown in Fig.\,\ref{figure_1}, the best-fitting FDM particle mass becomes $m=1.1\times 10^{-23}$ eV with $\chi^2\simeq 27.5$. Similarly, if we switch to using the stellar velocity dispersion $\sigma_\star$ at $R_{1/2}=1$ kpc, the best fitting mass remains at $m=1.2\times 10^{-23}$ eV, but with $\chi^2\simeq 14.4$. Therefore, the impact of ignoring the radius dependence of $\sigma$ and $\sigma_\star$ is small. 

Based on the preceding discussion, it is imperative to assess the influence of the parameter $k$, while the impact of $r_s$ can be disregarded. To investigate this, we vary $k$ from $2.7$ to $3.5$ in increments of $0.1$. Our findings reveal that the optimal $\chi^2$ initially decreases and subsequently increases as $k$ is incremented, reaching a minimum of $\chi^2\simeq 12.5$ at $k=2.8$. However, the maximum value of $\chi^2$ exceeds $100$ when $k=3.5$, indicating a poor fit. Additionally, the corresponding best-fitting FDM particle mass varies within the approximate range of $1-2\times 10^{-23}$ eV. The optimal FDM particle mass is determined to be $m=1.8\times 10^{-23}$ eV when $k=2.8$, and the corresponding profile is depicted in the right panel of Fig.\,\ref{figure_3}.

In the aforementioned analysis, we have fixed four observational parameters of the galaxy: the dynamical mass within $3R_e$, age, 2D half-light radius, and total stellar mass.
Here we discuss the impact of these parameters on the inferred particle mass. By maintaining $k$ and $r_s$ at their default values, we systematically vary each of the four parameters within the range of uncertainties provided by \cite{montes2024almost}, while holding the other three constant.
Our results reveal that the inferred particle mass varies within the ranges of $[1.1,1.4]\times 10^{-23}$ eV, $[1.0,1.3]\times 10^{-23}$ eV, $[1.1,1.2]\times 10^{-23}$ eV, and $[0.9,1.6]\times 10^{-23}$ eV for the respective parameters. Notably, the $\chi^2$ value peaks at 46.9 when the total stellar mass $M_\star$ is varied, while in the other cases, it remains below approximately 24. This discrepancy can be attributed to the direct relationship between $M_\star$ and the 2D stellar surface density distribution. Specifically, adjusting $M_\star$ to a smaller (larger) value tends to shift the fitting results towards the lower (upper) boundaries of the surface density. 

The impact of the uncertainty in the stellar mass $M_\star$ compared to the scale  $a_i$ in the range of $[0.5,1.5]$ kpc on the goodness of fit is notable. While both $M_\star$ and $a_i$ are solely present in the initial stellar density profile in Eq.\,\ref{Plum}, variations in $M_\star$ significantly affect the fit quality, whereas changes in $a_i$ have minimal impact. This discrepancy can be attributed to the fact that $M_\star$ determines the normalization of the initial density profile, while $a_i$ also influences its shape. For instance, let us consider two values $a_{i,1}$ and $a_{i,2}$ with $a_{i,1}<a_{i,2}$. The initial profile for $a_{i,2}$ exhibits lower density in the inner region and higher density in the outer region compared to $a_{i,1}$. The profile corresponding to $a_{i,2}$ can be viewed as a result equivalent to that stemming from the evolutionary process of an initial state characterized by $a_{i,1}$, even though there is no direct temporal connection between the two profiles. As the heating rate diminishes over time, the two profiles would eventually converge to the same profile after a sufficiently long period. Therefore, the value of $a_i$ does not significantly affect the final results.

\section{Discussions\label{discussions}}
\subsection{Comparison of the FDM particle mass with other studies}
Following our previous discussion, it is evident that a FDM particle mass within the range of approximately $m\simeq 1-2\times 10^{-23}$ eV can account for the observed stellar distribution in Nube. However, one limitation of our analysis is the lack of consideration for the impact of baryonic matter. Although the current surface stellar density is low enough to disregard baryonic effects, the density at the time of Nube's formation was higher and may have influenced the expansion of the stellar distribution due to star formation feedback. As a result, the actual FDM mass required may be slightly larger than our obtained value. Nevertheless, incorporating baryonic influence would necessitate further development of theoretical descriptions and simulations in the future. Therefore, for the time being, we temporarily overlook the impact of baryons.

There are several other studies which derived  FDM
mass similar to our result. For example,  Lora et al. demonstrated that a FDM mass within the range of $0.3\times 10^{-22}$ eV $<m<1\times 10^{-22}$ eV can adress the problem of
the longevity of the cold clump in Ursa Minor and the problem of the rapid orbital decay of the globular clusters (GCs) in Fornax \cite{Lora_2012}. Hložek et al. derived a lower limit of $m\gtrsim 10^{-24}$ eV based on the cosmic microwave background data from Planck \cite{Hlo_ek_2018}. Gonz\'{a}lez-Moralesthe et al. found an upper limit $m<0.4\times 10^{-22}$ eV at 97.5$\%$ confidence level by fitting to the $\langle \sigma^2_\text{los}\rangle$ data of Fornax and Sculptor \cite{Gonz_lez_Morales_2017}. Furthermore, Chiang et al. determined, based on an analysis of the dynamical heating effect of FDM on the Milky Way disk, that a FDM particle mass range of $m\simeq 0.5-0.7\times 10^{-22}$ eV is favored by the observed thick disc kinematics \cite{Chiang_2022}. In a more recent study, Bañares-Hernández et al. obtained a FDM particle mass of $m\simeq 2\times 10^{-23}$ eV \cite{Ba_ares_Hern_ndez_2023} by fitting to the rotation curves of nearby dwarf irregular galaxies.

There are also some studies indicating even lower FDM particle mass. For instance, Bernal et al. derived a FDM particle mass of $m=0.554\times 10^{-23}$ eV by fitting the soliton$+$NFW profile to the observed rotation curves of 18 high resolution low surface brightness galaxies \cite{Bernal_2017}. Additionally, Paredes and Michinel demonstrated that the quantum pressure of FDM with a particle mass of $m=2\times 10^{-24}$ eV can naturally account for the relative displacement between dark and ordinary matter in the galactic cluster Abell 3827, representing empirical evidence of dark matter interactions beyond gravity \cite{Paredes_2016}.

However, numerous strong constraints have emerged from studies involving the Lyman-$\alpha$ forest, subhalo mass function, and dynamical heating effect in dwarf galaxie, which yield values significantly higher than our findings. For instance, analyses of high-resolution observational data of the Lyman-$\alpha$ forest have led to strong lower bounds in the range of $m\gtrsim 0.7-20\times 10^{-21}$ \cite{Armengaud_2017,Ir_i__2017,Kobayashi_2017,Nori_2018,Rogers_2021}. However, it is important to note that these constraints are subject to numerous sources of uncertainties, such as astrophysical assumptions and data interpretation uncertainties \cite{Chiang_2022}. Additionally, the quantum pressure in FDM will suppress the FDM subhalo abundance in the low-mass tail, resulting in lower bounds on the FDM particle mass in the range of $m\gtrsim 2-5\times 10^{-21}$ eV \cite{Nadler_2019,Nadler_2021,Benito_2020,Schutz_2020}. However, it is worth noting that the subhalo mass functions obtained from different approaches themselves conflict \cite{Chiang_2022}. Furthermore, Marsh and Niemeyer obtained a lower bound of $m\gtrsim 0.6-1\times 10^{-19}$ eV by analysing the dynamical heating effect on the half-light radius of the star cluster at the center of Eridanus II \cite{Marsh_2019}. Nevertheless, Chiang et al. have argued that the soliton oscillations period in Eridanus II is too long to significantly heat the stars \cite{Chiang_2021}. Similarly, Dalal and Kravtsov utilized the sizes and stellar radial velocities data of Segue 1 and Segue 2 to establish a lower limit on the FDM particle mass of $m>3\times 10^{-19}$ eV. However, Dutta Chowdhury et al. have argued that there is a need to consider the tidal mass loss of these two dwarf galaxies as they are satellites of the Milky Way. It is important to acknowledge that many other constraints are in tension with our findings, but it is crucial to emphasize that most of them are the subject of significant debate. More details about the debates can be found in Refs. \cite{Chiang_2022,Ba_ares_Hern_ndez_2023,Ferreira_2021}.

Hence, the constraints on the mass of FDM particles derived from various studies display notable discrepancies. These differences can be attributed, in part, to errors in the observational data, uncertainties in astrophysical models, and the current limitations in our comprehension of FDM. Consequently, it is imperative to undertake future theoretical research on FDM and advance simulation methodologies to corroborate and refine these findings.

\subsection{Is Nube a specific field UDG?}
In this paper, we propose a novel mechanism utilizing FDM dynamical heating effect to explain the large extension of the stellar distribution of Nube. It is conceivable that this mechanism may also be applicable in elucidating the formation of UDGs. However, it is imperative to meticulously verify that this mechanism is consistent with existing observations, given that classical dwarf galaxies and most UDGs do not exhibit the same extent as Nube. We emphasize that the heating effect of FDM is only significant when the halo is not within a host halo \footnote{The projected distance between Nube and its likely host halo UGC 929 is about $435$ kpc. Additionally, other observational indicators suggest that Nube has not experienced significant tidal forces \cite{montes2024almost}.} as the tidal strip can result in a significant suppression of this effect \cite{Schive_2020}. Therefore, in the subsequent discussions, we concentrate on field UDGs and isolated classical dwarf galaxies.

To begin with, numerous observations revealed that most of the field UDGs are blue, HI-rich, star-forming, and with irregular morphologies, these features indicate that their ages are young \cite{leisman2017almost,Prole_2019}. In a more quantitative analysis \cite{Rom_n_2017}, Rom\'{a}n and Trujillo discovered that the ages of their sample of blue UDGs, located at a projected distance of more than 200 kpc from their host group central regions, are approximately less than 1 Gyr. Therefore, the stars in field UDGs cannot have been heated for as long as Nube (10.2 Gyr), and the formation of stars continues in these systems. These two factors limit their 2D half-light radius to only expand to $1.5-5$ kpc, which is smaller than Nube. Consequently, we propose that Nube may be considered a specific field UDG, not accreted into a host halo to become a cluster UDG, but remaining in the field for a sufficiently long period. Additionally, some observations have found that field UDGs typically have few or no GCs \cite{Jones_2022}, similar to classical dwarf galaxies. Similarly, no GCs associated with Nube have been detected \cite{montes2024almost}, indicating another similarity between Nube and field UDGs.

The absence of GCs in field UDGs, similar to that found in dwarf galaxies, suggests that the formation mechanism of field UDGs does not influence the formation of GCs \cite{Jones_2022}. However, some hydrodynamic simulations have indicated that the number of GCs should be greater than the observed results within the framework of CDM \cite{Benavides_2023,Jones_2022}. If we consider that UDGs are also formed through the dynamical heating effect of FDM, then the number of GCs obviously will not be affected.  Furthermore, although field UDGs are typically HI-rich, many studies have found that they still adhere to the standard HI size-radius relations \cite{Wang_2016,leisman2017almost,Gault_2021}. This property implies that their formation mechanism does not affect the HI radius. As the HI distribution is generally much more extended than the stars, the FDM heating effect can be disregarded in the nearby region of the HI radius due to the low FDM density there, as shown in Fig.\,\ref{figure_2}. Therefore, the FDM heating mechanism has almost no impact on the HI size-mass relations.

Similarly, the majority of the isolated classical dwarf galaxies exhibit ongoing star formation, indicating that their stellar distribution is not in direct conflict with the FDM heating effect. Notably, many dwarf galaxies display a stellar age gradient in their stellar distribution \cite{Pucha_2019,Higgs_2021}, with older stars being more extended than younger stars. This observation also aligns with the dynamical heating effect, as the heating time for older stars is longer than for younger stars. 

Finally, we focus on a recently observed dwarf galaxy in the nearby universe \cite{carleton2024pearls}, which may potentially conflict with the aforementioned mechanism. This galaxy is an isolated quiescent galaxy observed in the JWST PEARLS program, with a very red optical spectrum indicating an old stellar population, and its star formation rate is $4\times 10^{-4} M_\odot$ yr$^{-1}$. Note that its half-light radius is only $\sim 0.53$ kpc, which may conflict with the FDM heating effect. However, in \cite{carleton2024pearls}, the authors point out that they cannot rule out the possibility of past interactions between this galaxy and other galaxies, such as J1227, which could influence the stellar and dark matter structures, thus rendering the aforementioned conflict ambiguous.

In summary, our discussions suggest that Nube may be considered as a specific field of UDGs, and the FDM heating effect may also be applicable in elucidating the formation of UDGs. However, in order to account for the irregular morphologies of field UDGs, the baryonic influence must be considered, which is contingent on future studies. Our arguments above are primarily qualitative, and further quantitative assessments are necessary to establish the feasibility of this perspective.

\section{Conclusion\label{conclusion}}
In this study, we present a novel mechanism that leverages the FDM dynamical heating effect to elucidate the stellar distribution anomaly observed in Nube. Our findings indicate that a FDM particle mass within the range of $m\simeq 1-2\times 10^{-23}$ eV can naturally explain the extensive extension of the stellar distribution. While our results align with certain previous studies, they also face tension with some constraints documented in the literature. It is noteworthy that many of these constraints are accompanied by substantial uncertainties and are the subject of ongoing debate. 

Furthermore, we observe that the FDM dynamical heating effect may offer insights into understanding the formation of field UDGs and qualitatively aligns with observational outcomes. From this perspective, Nube could be considered a distinctive field UDG, having not been accreted into a host halo to become a cluster UDG, but instead persisting in the field for a long duration. However, it is important to emphasize that our demonstration of the self-consistency of this viewpoint is primarily qualitative. Future research endeavors should focus on quantitative assessments to establish the feasibility of this perspective.

\acknowledgments
We thank Run-Min Yao for helpful discussions and verification of the calculation results. This work is supported by the National Natural Science Foundation of China under grant No. 12175248.


\bibliography{FDM}

\begin{thebibliography}{89}%
\makeatletter
\providecommand \@ifxundefined [1]{%
 \@ifx{#1\undefined}
}%
\providecommand \@ifnum [1]{%
 \ifnum #1\expandafter \@firstoftwo
 \else \expandafter \@secondoftwo
 \fi
}%
\providecommand \@ifx [1]{%
 \ifx #1\expandafter \@firstoftwo
 \else \expandafter \@secondoftwo
 \fi
}%
\providecommand \natexlab [1]{#1}%
\providecommand \enquote  [1]{``#1''}%
\providecommand \bibnamefont  [1]{#1}%
\providecommand \bibfnamefont [1]{#1}%
\providecommand \citenamefont [1]{#1}%
\providecommand \href@noop [0]{\@secondoftwo}%
\providecommand \href [0]{\begingroup \@sanitize@url \@href}%
\providecommand \@href[1]{\@@startlink{#1}\@@href}%
\providecommand \@@href[1]{\endgroup#1\@@endlink}%
\providecommand \@sanitize@url [0]{\catcode `\\12\catcode `\$12\catcode `\&12\catcode `\#12\catcode `\^12\catcode `\_12\catcode `\%12\relax}%
\providecommand \@@startlink[1]{}%
\providecommand \@@endlink[0]{}%
\providecommand \url  [0]{\begingroup\@sanitize@url \@url }%
\providecommand \@url [1]{\endgroup\@href {#1}{\urlprefix }}%
\providecommand \urlprefix  [0]{URL }%
\providecommand \Eprint [0]{\href }%
\providecommand \doibase [0]{https://doi.org/}%
\providecommand \selectlanguage [0]{\@gobble}%
\providecommand \bibinfo  [0]{\@secondoftwo}%
\providecommand \bibfield  [0]{\@secondoftwo}%
\providecommand \translation [1]{[#1]}%
\providecommand \BibitemOpen [0]{}%
\providecommand \bibitemStop [0]{}%
\providecommand \bibitemNoStop [0]{.\EOS\space}%
\providecommand \EOS [0]{\spacefactor3000\relax}%
\providecommand \BibitemShut  [1]{\csname bibitem#1\endcsname}%
\let\auto@bib@innerbib\@empty
\bibitem [{\citenamefont {{Zwicky}}(1933)}]{zwicky6spectral}%
  \BibitemOpen
  \bibfield  {author} {\bibinfo {author} {\bibfnamefont {F.}~\bibnamefont {{Zwicky}}},\ }\bibfield  {title} {\bibinfo {title} {{Die Rotverschiebung von extragalaktischen Nebeln}},\ }\href@noop {} {\bibfield  {journal} {\bibinfo  {journal} {Helvetica Physica Acta}\ }\textbf {\bibinfo {volume} {6}},\ \bibinfo {pages} {110} (\bibinfo {year} {1933})}\BibitemShut {NoStop}%
\bibitem [{\citenamefont {{Rubin}}\ and\ \citenamefont {{Ford}}(1970)}]{rubin1970rotation}%
  \BibitemOpen
  \bibfield  {author} {\bibinfo {author} {\bibfnamefont {V.~C.}\ \bibnamefont {{Rubin}}}\ and\ \bibinfo {author} {\bibfnamefont {J.}~\bibnamefont {{Ford}}, \bibfnamefont {W.~Kent}},\ }\bibfield  {title} {\bibinfo {title} {{Rotation of the Andromeda Nebula from a Spectroscopic Survey of Emission Regions}},\ }\href {https://doi.org/10.1086/150317} {\bibfield  {journal} {\bibinfo  {journal} {\apj}\ }\textbf {\bibinfo {volume} {159}},\ \bibinfo {pages} {379} (\bibinfo {year} {1970})}\BibitemShut {NoStop}%
\bibitem [{\citenamefont {{Flores}}\ and\ \citenamefont {{Primack}}(1994)}]{flores1994observational}%
  \BibitemOpen
  \bibfield  {author} {\bibinfo {author} {\bibfnamefont {R.~A.}\ \bibnamefont {{Flores}}}\ and\ \bibinfo {author} {\bibfnamefont {J.~R.}\ \bibnamefont {{Primack}}},\ }\bibfield  {title} {\bibinfo {title} {{Observational and Theoretical Constraints on Singular Dark Matter Halos}},\ }\href {https://doi.org/10.1086/187350} {\bibfield  {journal} {\bibinfo  {journal} {\apj}\ }\textbf {\bibinfo {volume} {427}},\ \bibinfo {pages} {L1} (\bibinfo {year} {1994})},\ \Eprint {https://arxiv.org/abs/astro-ph/9402004} {arXiv:astro-ph/9402004 [astro-ph]} \BibitemShut {NoStop}%
\bibitem [{\citenamefont {{Moore}}(1994)}]{moore1994evidence}%
  \BibitemOpen
  \bibfield  {author} {\bibinfo {author} {\bibfnamefont {B.}~\bibnamefont {{Moore}}},\ }\bibfield  {title} {\bibinfo {title} {{Evidence against dissipation-less dark matter from observations of galaxy haloes}},\ }\href {https://doi.org/10.1038/370629a0} {\bibfield  {journal} {\bibinfo  {journal} {\nat}\ }\textbf {\bibinfo {volume} {370}},\ \bibinfo {pages} {629} (\bibinfo {year} {1994})}\BibitemShut {NoStop}%
\bibitem [{\citenamefont {de~Blok}(2010)}]{de2010core}%
  \BibitemOpen
  \bibfield  {author} {\bibinfo {author} {\bibfnamefont {W.~J.~G.}\ \bibnamefont {de~Blok}},\ }\bibfield  {title} {\bibinfo {title} {{The Core-Cusp Problem}},\ }\href {https://doi.org/10.1155/2010/789293} {\bibfield  {journal} {\bibinfo  {journal} {Advances in Astronomy}\ }\textbf {\bibinfo {volume} {2010}},\ \bibinfo {pages} {1–14} (\bibinfo {year} {2010})}\BibitemShut {NoStop}%
\bibitem [{\citenamefont {Moore}\ \emph {et~al.}(1999)\citenamefont {Moore}, \citenamefont {Ghigna}, \citenamefont {Governato}, \citenamefont {Lake}, \citenamefont {Quinn}, \citenamefont {Stadel},\ and\ \citenamefont {Tozzi}}]{moore1999dark}%
  \BibitemOpen
  \bibfield  {author} {\bibinfo {author} {\bibfnamefont {B.}~\bibnamefont {Moore}}, \bibinfo {author} {\bibfnamefont {S.}~\bibnamefont {Ghigna}}, \bibinfo {author} {\bibfnamefont {F.}~\bibnamefont {Governato}}, \bibinfo {author} {\bibfnamefont {G.}~\bibnamefont {Lake}}, \bibinfo {author} {\bibfnamefont {T.}~\bibnamefont {Quinn}}, \bibinfo {author} {\bibfnamefont {J.}~\bibnamefont {Stadel}},\ and\ \bibinfo {author} {\bibfnamefont {P.}~\bibnamefont {Tozzi}},\ }\bibfield  {title} {\bibinfo {title} {{Dark Matter Substructure within Galactic Halos}},\ }\href {https://doi.org/10.1086/312287} {\bibfield  {journal} {\bibinfo  {journal} {The Astrophysical Journal}\ }\textbf {\bibinfo {volume} {524}},\ \bibinfo {pages} {L19–L22} (\bibinfo {year} {1999})}\BibitemShut {NoStop}%
\bibitem [{\citenamefont {Klypin}\ \emph {et~al.}(1999)\citenamefont {Klypin}, \citenamefont {Kravtsov}, \citenamefont {Valenzuela},\ and\ \citenamefont {Prada}}]{klypin1999missing}%
  \BibitemOpen
  \bibfield  {author} {\bibinfo {author} {\bibfnamefont {A.}~\bibnamefont {Klypin}}, \bibinfo {author} {\bibfnamefont {A.~V.}\ \bibnamefont {Kravtsov}}, \bibinfo {author} {\bibfnamefont {O.}~\bibnamefont {Valenzuela}},\ and\ \bibinfo {author} {\bibfnamefont {F.}~\bibnamefont {Prada}},\ }\bibfield  {title} {\bibinfo {title} {{Where Are the Missing Galactic Satellites?}},\ }\href {https://doi.org/10.1086/307643} {\bibfield  {journal} {\bibinfo  {journal} {The Astrophysical Journal}\ }\textbf {\bibinfo {volume} {522}},\ \bibinfo {pages} {82–92} (\bibinfo {year} {1999})}\BibitemShut {NoStop}%
\bibitem [{\citenamefont {Zavala}\ \emph {et~al.}(2009)\citenamefont {Zavala}, \citenamefont {Jing}, \citenamefont {Faltenbacher}, \citenamefont {Yepes}, \citenamefont {Hoffman}, \citenamefont {Gottlöber},\ and\ \citenamefont {Catinella}}]{zavala2009velocity}%
  \BibitemOpen
  \bibfield  {author} {\bibinfo {author} {\bibfnamefont {J.}~\bibnamefont {Zavala}}, \bibinfo {author} {\bibfnamefont {Y.~P.}\ \bibnamefont {Jing}}, \bibinfo {author} {\bibfnamefont {A.}~\bibnamefont {Faltenbacher}}, \bibinfo {author} {\bibfnamefont {G.}~\bibnamefont {Yepes}}, \bibinfo {author} {\bibfnamefont {Y.}~\bibnamefont {Hoffman}}, \bibinfo {author} {\bibfnamefont {S.}~\bibnamefont {Gottlöber}},\ and\ \bibinfo {author} {\bibfnamefont {B.}~\bibnamefont {Catinella}},\ }\bibfield  {title} {\bibinfo {title} {The velocity function in the local environment from {$\Lambda$CDM} and {$\Lambda$WDM} constrained simulations},\ }\href {https://doi.org/10.1088/0004-637x/700/2/1779} {\bibfield  {journal} {\bibinfo  {journal} {The Astrophysical Journal}\ }\textbf {\bibinfo {volume} {700}},\ \bibinfo {pages} {1779–1793} (\bibinfo {year} {2009})}\BibitemShut {NoStop}%
\bibitem [{\citenamefont {Boylan-Kolchin}\ \emph {et~al.}(2011)\citenamefont {Boylan-Kolchin}, \citenamefont {Bullock},\ and\ \citenamefont {Kaplinghat}}]{boylan2011too}%
  \BibitemOpen
  \bibfield  {author} {\bibinfo {author} {\bibfnamefont {M.}~\bibnamefont {Boylan-Kolchin}}, \bibinfo {author} {\bibfnamefont {J.~S.}\ \bibnamefont {Bullock}},\ and\ \bibinfo {author} {\bibfnamefont {M.}~\bibnamefont {Kaplinghat}},\ }\bibfield  {title} {\bibinfo {title} {{Too big to fail? The puzzling darkness of massive Milky Way subhaloes}},\ }\href {https://doi.org/10.1111/j.1745-3933.2011.01074.x} {\bibfield  {journal} {\bibinfo  {journal} {Monthly Notices of the Royal Astronomical Society: Letters}\ }\textbf {\bibinfo {volume} {415}},\ \bibinfo {pages} {L40–L44} (\bibinfo {year} {2011})}\BibitemShut {NoStop}%
\bibitem [{\citenamefont {Boylan-Kolchin}\ \emph {et~al.}(2012)\citenamefont {Boylan-Kolchin}, \citenamefont {Bullock},\ and\ \citenamefont {Kaplinghat}}]{boylan2012milky}%
  \BibitemOpen
  \bibfield  {author} {\bibinfo {author} {\bibfnamefont {M.}~\bibnamefont {Boylan-Kolchin}}, \bibinfo {author} {\bibfnamefont {J.~S.}\ \bibnamefont {Bullock}},\ and\ \bibinfo {author} {\bibfnamefont {M.}~\bibnamefont {Kaplinghat}},\ }\bibfield  {title} {\bibinfo {title} {{The Milky Way’s bright satellites as an apparent failure of {$\Lambda$CDM}}},\ }\href {https://doi.org/10.1111/j.1365-2966.2012.20695.x} {\bibfield  {journal} {\bibinfo  {journal} {Monthly Notices of the Royal Astronomical Society}\ }\textbf {\bibinfo {volume} {422}},\ \bibinfo {pages} {1203–1218} (\bibinfo {year} {2012})}\BibitemShut {NoStop}%
\bibitem [{\citenamefont {Tollerud}\ \emph {et~al.}(2014)\citenamefont {Tollerud}, \citenamefont {Boylan-Kolchin},\ and\ \citenamefont {Bullock}}]{tollerud2014m31}%
  \BibitemOpen
  \bibfield  {author} {\bibinfo {author} {\bibfnamefont {E.~J.}\ \bibnamefont {Tollerud}}, \bibinfo {author} {\bibfnamefont {M.}~\bibnamefont {Boylan-Kolchin}},\ and\ \bibinfo {author} {\bibfnamefont {J.~S.}\ \bibnamefont {Bullock}},\ }\bibfield  {title} {\bibinfo {title} {{M31 satellite masses compared to {$\Lambda$CDM} subhaloes}},\ }\href {https://doi.org/10.1093/mnras/stu474} {\bibfield  {journal} {\bibinfo  {journal} {Monthly Notices of the Royal Astronomical Society}\ }\textbf {\bibinfo {volume} {440}},\ \bibinfo {pages} {3511–3519} (\bibinfo {year} {2014})}\BibitemShut {NoStop}%
\bibitem [{\citenamefont {Garrison-Kimmel}\ \emph {et~al.}(2014)\citenamefont {Garrison-Kimmel}, \citenamefont {Boylan-Kolchin}, \citenamefont {Bullock},\ and\ \citenamefont {Kirby}}]{garrison2014too}%
  \BibitemOpen
  \bibfield  {author} {\bibinfo {author} {\bibfnamefont {S.}~\bibnamefont {Garrison-Kimmel}}, \bibinfo {author} {\bibfnamefont {M.}~\bibnamefont {Boylan-Kolchin}}, \bibinfo {author} {\bibfnamefont {J.~S.}\ \bibnamefont {Bullock}},\ and\ \bibinfo {author} {\bibfnamefont {E.~N.}\ \bibnamefont {Kirby}},\ }\bibfield  {title} {\bibinfo {title} {{Too big to fail in the Local Group}},\ }\href {https://doi.org/10.1093/mnras/stu1477} {\bibfield  {journal} {\bibinfo  {journal} {Monthly Notices of the Royal Astronomical Society}\ }\textbf {\bibinfo {volume} {444}},\ \bibinfo {pages} {222–236} (\bibinfo {year} {2014})}\BibitemShut {NoStop}%
\bibitem [{\citenamefont {Bode}\ \emph {et~al.}(2001)\citenamefont {Bode}, \citenamefont {Ostriker},\ and\ \citenamefont {Turok}}]{bode2001halo}%
  \BibitemOpen
  \bibfield  {author} {\bibinfo {author} {\bibfnamefont {P.}~\bibnamefont {Bode}}, \bibinfo {author} {\bibfnamefont {J.~P.}\ \bibnamefont {Ostriker}},\ and\ \bibinfo {author} {\bibfnamefont {N.}~\bibnamefont {Turok}},\ }\bibfield  {title} {\bibinfo {title} {{Halo Formation in Warm Dark Matter Models}},\ }\href {https://doi.org/10.1086/321541} {\bibfield  {journal} {\bibinfo  {journal} {The Astrophysical Journal}\ }\textbf {\bibinfo {volume} {556}},\ \bibinfo {pages} {93–107} (\bibinfo {year} {2001})}\BibitemShut {NoStop}%
\bibitem [{\citenamefont {Schneider}\ \emph {et~al.}(2017)\citenamefont {Schneider}, \citenamefont {Trujillo-Gomez}, \citenamefont {Papastergis}, \citenamefont {Reed},\ and\ \citenamefont {Lake}}]{schneider2017hints}%
  \BibitemOpen
  \bibfield  {author} {\bibinfo {author} {\bibfnamefont {A.}~\bibnamefont {Schneider}}, \bibinfo {author} {\bibfnamefont {S.}~\bibnamefont {Trujillo-Gomez}}, \bibinfo {author} {\bibfnamefont {E.}~\bibnamefont {Papastergis}}, \bibinfo {author} {\bibfnamefont {D.~S.}\ \bibnamefont {Reed}},\ and\ \bibinfo {author} {\bibfnamefont {G.}~\bibnamefont {Lake}},\ }\bibfield  {title} {\bibinfo {title} {{Hints against the cold and collisionless nature of dark matter from the galaxy velocity function}},\ }\href {https://doi.org/10.1093/mnras/stx1294} {\bibfield  {journal} {\bibinfo  {journal} {Monthly Notices of the Royal Astronomical Society}\ }\textbf {\bibinfo {volume} {470}},\ \bibinfo {pages} {1542–1558} (\bibinfo {year} {2017})}\BibitemShut {NoStop}%
\bibitem [{\citenamefont {Lovell}\ \emph {et~al.}(2017)\citenamefont {Lovell}, \citenamefont {Gonzalez-Perez}, \citenamefont {Bose}, \citenamefont {Boyarsky}, \citenamefont {Cole}, \citenamefont {Frenk},\ and\ \citenamefont {Ruchayskiy}}]{lovell2017addressing}%
  \BibitemOpen
  \bibfield  {author} {\bibinfo {author} {\bibfnamefont {M.~R.}\ \bibnamefont {Lovell}}, \bibinfo {author} {\bibfnamefont {V.}~\bibnamefont {Gonzalez-Perez}}, \bibinfo {author} {\bibfnamefont {S.}~\bibnamefont {Bose}}, \bibinfo {author} {\bibfnamefont {A.}~\bibnamefont {Boyarsky}}, \bibinfo {author} {\bibfnamefont {S.}~\bibnamefont {Cole}}, \bibinfo {author} {\bibfnamefont {C.~S.}\ \bibnamefont {Frenk}},\ and\ \bibinfo {author} {\bibfnamefont {O.}~\bibnamefont {Ruchayskiy}},\ }\bibfield  {title} {\bibinfo {title} {Addressing the too big to fail problem with baryon physics and sterile neutrino dark matter},\ }\href {https://doi.org/10.1093/mnras/stx621} {\bibfield  {journal} {\bibinfo  {journal} {Monthly Notices of the Royal Astronomical Society}\ }\textbf {\bibinfo {volume} {468}},\ \bibinfo {pages} {2836–2849} (\bibinfo {year} {2017})}\BibitemShut {NoStop}%
\bibitem [{\citenamefont {Spergel}\ and\ \citenamefont {Steinhardt}(2000)}]{spergel2000observational}%
  \BibitemOpen
  \bibfield  {author} {\bibinfo {author} {\bibfnamefont {D.~N.}\ \bibnamefont {Spergel}}\ and\ \bibinfo {author} {\bibfnamefont {P.~J.}\ \bibnamefont {Steinhardt}},\ }\bibfield  {title} {\bibinfo {title} {{Observational Evidence for Self-Interacting Cold Dark Matter}},\ }\href {https://doi.org/10.1103/physrevlett.84.3760} {\bibfield  {journal} {\bibinfo  {journal} {Physical Review Letters}\ }\textbf {\bibinfo {volume} {84}},\ \bibinfo {pages} {3760–3763} (\bibinfo {year} {2000})}\BibitemShut {NoStop}%
\bibitem [{\citenamefont {Tulin}\ and\ \citenamefont {Yu}(2018)}]{tulin2018dark}%
  \BibitemOpen
  \bibfield  {author} {\bibinfo {author} {\bibfnamefont {S.}~\bibnamefont {Tulin}}\ and\ \bibinfo {author} {\bibfnamefont {H.-B.}\ \bibnamefont {Yu}},\ }\bibfield  {title} {\bibinfo {title} {Dark matter self-interactions and small scale structure},\ }\href {https://doi.org/10.1016/j.physrep.2017.11.004} {\bibfield  {journal} {\bibinfo  {journal} {Physics Reports}\ }\textbf {\bibinfo {volume} {730}},\ \bibinfo {pages} {1–57} (\bibinfo {year} {2018})}\BibitemShut {NoStop}%
\bibitem [{\citenamefont {Adhikari}\ \emph {et~al.}(2022)\citenamefont {Adhikari} \emph {et~al.}}]{adhikari2022astrophysical}%
  \BibitemOpen
  \bibfield  {author} {\bibinfo {author} {\bibfnamefont {S.}~\bibnamefont {Adhikari}} \emph {et~al.},\ }\href@noop {} {\bibinfo {title} {{Astrophysical Tests of Dark Matter Self-Interactions}}} (\bibinfo {year} {2022}),\ \Eprint {https://arxiv.org/abs/2207.10638} {arXiv:2207.10638 [astro-ph.CO]} \BibitemShut {NoStop}%
\bibitem [{\citenamefont {Hu}\ \emph {et~al.}(2000)\citenamefont {Hu}, \citenamefont {Barkana},\ and\ \citenamefont {Gruzinov}}]{hu2000fuzzy}%
  \BibitemOpen
  \bibfield  {author} {\bibinfo {author} {\bibfnamefont {W.}~\bibnamefont {Hu}}, \bibinfo {author} {\bibfnamefont {R.}~\bibnamefont {Barkana}},\ and\ \bibinfo {author} {\bibfnamefont {A.}~\bibnamefont {Gruzinov}},\ }\bibfield  {title} {\bibinfo {title} {{Fuzzy Cold Dark Matter: The Wave Properties of Ultralight Particles}},\ }\href {https://doi.org/10.1103/physrevlett.85.1158} {\bibfield  {journal} {\bibinfo  {journal} {Physical Review Letters}\ }\textbf {\bibinfo {volume} {85}},\ \bibinfo {pages} {1158–1161} (\bibinfo {year} {2000})}\BibitemShut {NoStop}%
\bibitem [{\citenamefont {{Peebles}}(2000)}]{peebles2000fluid}%
  \BibitemOpen
  \bibfield  {author} {\bibinfo {author} {\bibfnamefont {P.~J.~E.}\ \bibnamefont {{Peebles}}},\ }\bibfield  {title} {\bibinfo {title} {{Fluid Dark Matter}},\ }\href {https://doi.org/10.1086/312677} {\bibfield  {journal} {\bibinfo  {journal} {\apj}\ }\textbf {\bibinfo {volume} {534}},\ \bibinfo {pages} {L127} (\bibinfo {year} {2000})},\ \Eprint {https://arxiv.org/abs/astro-ph/0002495} {arXiv:astro-ph/0002495 [astro-ph]} \BibitemShut {NoStop}%
\bibitem [{\citenamefont {Hui}\ \emph {et~al.}(2017)\citenamefont {Hui}, \citenamefont {Ostriker}, \citenamefont {Tremaine},\ and\ \citenamefont {Witten}}]{hui2017ultralight}%
  \BibitemOpen
  \bibfield  {author} {\bibinfo {author} {\bibfnamefont {L.}~\bibnamefont {Hui}}, \bibinfo {author} {\bibfnamefont {J.~P.}\ \bibnamefont {Ostriker}}, \bibinfo {author} {\bibfnamefont {S.}~\bibnamefont {Tremaine}},\ and\ \bibinfo {author} {\bibfnamefont {E.}~\bibnamefont {Witten}},\ }\bibfield  {title} {\bibinfo {title} {Ultralight scalars as cosmological dark matter},\ }\bibfield  {journal} {\bibinfo  {journal} {Physical Review D}\ }\textbf {\bibinfo {volume} {95}},\ \href {https://doi.org/10.1103/physrevd.95.043541} {10.1103/physrevd.95.043541} (\bibinfo {year} {2017})\BibitemShut {NoStop}%
\bibitem [{\citenamefont {Hui}(2021)}]{hui2021wave}%
  \BibitemOpen
  \bibfield  {author} {\bibinfo {author} {\bibfnamefont {L.}~\bibnamefont {Hui}},\ }\bibfield  {title} {\bibinfo {title} {{Wave Dark Matter}},\ }\href {https://doi.org/10.1146/annurev-astro-120920-010024} {\bibfield  {journal} {\bibinfo  {journal} {Annual Review of Astronomy and Astrophysics}\ }\textbf {\bibinfo {volume} {59}},\ \bibinfo {pages} {247–289} (\bibinfo {year} {2021})}\BibitemShut {NoStop}%
\bibitem [{\citenamefont {Matos}\ \emph {et~al.}(2023)\citenamefont {Matos}, \citenamefont {Ureña-López},\ and\ \citenamefont {Lee}}]{matos2023short}%
  \BibitemOpen
  \bibfield  {author} {\bibinfo {author} {\bibfnamefont {T.}~\bibnamefont {Matos}}, \bibinfo {author} {\bibfnamefont {L.~A.}\ \bibnamefont {Ureña-López}},\ and\ \bibinfo {author} {\bibfnamefont {J.-W.}\ \bibnamefont {Lee}},\ }\href@noop {} {\bibinfo {title} {{Short Review of the main achievements of the Scalar Field, Fuzzy, Ultralight, Wave, BEC Dark Matter model}}} (\bibinfo {year} {2023}),\ \Eprint {https://arxiv.org/abs/2312.00254} {arXiv:2312.00254 [astro-ph.CO]} \BibitemShut {NoStop}%
\bibitem [{\citenamefont {Khlopov}\ \emph {et~al.}(1999)\citenamefont {Khlopov}, \citenamefont {Sakharov},\ and\ \citenamefont {Sokoloff}}]{Khlopov_1999}%
  \BibitemOpen
  \bibfield  {author} {\bibinfo {author} {\bibfnamefont {M.}~\bibnamefont {Khlopov}}, \bibinfo {author} {\bibfnamefont {A.}~\bibnamefont {Sakharov}},\ and\ \bibinfo {author} {\bibfnamefont {D.}~\bibnamefont {Sokoloff}},\ }\bibfield  {title} {\bibinfo {title} {{The nonlinear modulation of the density distribution in standard axionic CDM and its cosmological impact}},\ }\href {https://doi.org/10.1016/s0920-5632(98)00511-8} {\bibfield  {journal} {\bibinfo  {journal} {Nuclear Physics B - Proceedings Supplements}\ }\textbf {\bibinfo {volume} {72}},\ \bibinfo {pages} {105–109} (\bibinfo {year} {1999})}\BibitemShut {NoStop}%
\bibitem [{\citenamefont {Janowiecki}\ \emph {et~al.}(2015)\citenamefont {Janowiecki} \emph {et~al.}}]{janowiecki2015almost}%
  \BibitemOpen
  \bibfield  {author} {\bibinfo {author} {\bibfnamefont {S.}~\bibnamefont {Janowiecki}} \emph {et~al.},\ }\bibfield  {title} {\bibinfo {title} {{(Almost)} dark {HI} sources in the {ALFALFA} survey: The intriguing case of {HI1232+20}},\ }\href {https://doi.org/10.1088/0004-637x/801/2/96} {\bibfield  {journal} {\bibinfo  {journal} {The Astrophysical Journal}\ }\textbf {\bibinfo {volume} {801}},\ \bibinfo {pages} {96} (\bibinfo {year} {2015})}\BibitemShut {NoStop}%
\bibitem [{\citenamefont {{Leisman}}\ \emph {et~al.}(2017)\citenamefont {{Leisman}} \emph {et~al.}}]{leisman2017almost}%
  \BibitemOpen
  \bibfield  {author} {\bibinfo {author} {\bibfnamefont {L.}~\bibnamefont {{Leisman}}} \emph {et~al.},\ }\bibfield  {title} {\bibinfo {title} {{(Almost) Dark Galaxies in the {ALFALFA} Survey: Isolated {HI}-bearing Ultra-diffuse Galaxies}},\ }\href {https://doi.org/10.3847/1538-4357/aa7575} {\bibfield  {journal} {\bibinfo  {journal} {\apj}\ }\textbf {\bibinfo {volume} {842}},\ \bibinfo {eid} {133} (\bibinfo {year} {2017})},\ \Eprint {https://arxiv.org/abs/1703.05293} {arXiv:1703.05293 [astro-ph.GA]} \BibitemShut {NoStop}%
\bibitem [{\citenamefont {Brunker}\ \emph {et~al.}(2019)\citenamefont {Brunker} \emph {et~al.}}]{brunker2019enigmatic}%
  \BibitemOpen
  \bibfield  {author} {\bibinfo {author} {\bibfnamefont {S.~W.}\ \bibnamefont {Brunker}} \emph {et~al.},\ }\bibfield  {title} {\bibinfo {title} {{The Enigmatic (Almost) Dark Galaxy {Coma P}: Distance Measurement and Stellar Populations from {HST} Imaging}},\ }\href {https://doi.org/10.3847/1538-3881/aafb39} {\bibfield  {journal} {\bibinfo  {journal} {The Astronomical Journal}\ }\textbf {\bibinfo {volume} {157}},\ \bibinfo {pages} {76} (\bibinfo {year} {2019})}\BibitemShut {NoStop}%
\bibitem [{\citenamefont {Montes}\ \emph {et~al.}(2024)\citenamefont {Montes} \emph {et~al.}}]{montes2024almost}%
  \BibitemOpen
  \bibfield  {author} {\bibinfo {author} {\bibfnamefont {M.}~\bibnamefont {Montes}} \emph {et~al.},\ }\bibfield  {title} {\bibinfo {title} {{An almost dark galaxy with the mass of the Small Magellanic Cloud}},\ }\href {https://doi.org/10.1051/0004-6361/202347667} {\bibfield  {journal} {\bibinfo  {journal} {Astronomy $\&$ Astrophysics}\ }\textbf {\bibinfo {volume} {681}},\ \bibinfo {pages} {A15} (\bibinfo {year} {2024})}\BibitemShut {NoStop}%
\bibitem [{\citenamefont {Tremaine}\ and\ \citenamefont {Gunn}(1979)}]{Tremaine:1979we}%
  \BibitemOpen
  \bibfield  {author} {\bibinfo {author} {\bibfnamefont {S.}~\bibnamefont {Tremaine}}\ and\ \bibinfo {author} {\bibfnamefont {J.~E.}\ \bibnamefont {Gunn}},\ }\bibfield  {title} {\bibinfo {title} {{Dynamical Role of Light Neutral Leptons in Cosmology}},\ }\href {https://doi.org/10.1103/PhysRevLett.42.407} {\bibfield  {journal} {\bibinfo  {journal} {Phys. Rev. Lett.}\ }\textbf {\bibinfo {volume} {42}},\ \bibinfo {pages} {407} (\bibinfo {year} {1979})}\BibitemShut {NoStop}%
\bibitem [{\citenamefont {Guth}\ \emph {et~al.}(2015)\citenamefont {Guth}, \citenamefont {Hertzberg},\ and\ \citenamefont {Prescod-Weinstein}}]{Guth:2014hsa}%
  \BibitemOpen
  \bibfield  {author} {\bibinfo {author} {\bibfnamefont {A.~H.}\ \bibnamefont {Guth}}, \bibinfo {author} {\bibfnamefont {M.~P.}\ \bibnamefont {Hertzberg}},\ and\ \bibinfo {author} {\bibfnamefont {C.}~\bibnamefont {Prescod-Weinstein}},\ }\bibfield  {title} {\bibinfo {title} {{Do Dark Matter Axions Form a Condensate with Long-Range Correlation?}},\ }\href {https://doi.org/10.1103/PhysRevD.92.103513} {\bibfield  {journal} {\bibinfo  {journal} {Phys. Rev. D}\ }\textbf {\bibinfo {volume} {92}},\ \bibinfo {pages} {103513} (\bibinfo {year} {2015})},\ \Eprint {https://arxiv.org/abs/1412.5930} {arXiv:1412.5930 [astro-ph.CO]} \BibitemShut {NoStop}%
\bibitem [{\citenamefont {Glauber}(1963)}]{article}%
  \BibitemOpen
  \bibfield  {author} {\bibinfo {author} {\bibfnamefont {R.}~\bibnamefont {Glauber}},\ }\bibfield  {title} {\bibinfo {title} {The quantum theory of optical coherence},\ }\href {https://doi.org/10.1103/PhysRev.130.2529} {\bibfield  {journal} {\bibinfo  {journal} {Phys. Rev.}\ }\textbf {\bibinfo {volume} {130}},\ \bibinfo {pages} {2529} (\bibinfo {year} {1963})}\BibitemShut {NoStop}%
\bibitem [{\citenamefont {Schive}\ \emph {et~al.}(2014{\natexlab{a}})\citenamefont {Schive}, \citenamefont {Chiueh},\ and\ \citenamefont {Broadhurst}}]{Schive_2014_1}%
  \BibitemOpen
  \bibfield  {author} {\bibinfo {author} {\bibfnamefont {H.-Y.}\ \bibnamefont {Schive}}, \bibinfo {author} {\bibfnamefont {T.}~\bibnamefont {Chiueh}},\ and\ \bibinfo {author} {\bibfnamefont {T.}~\bibnamefont {Broadhurst}},\ }\bibfield  {title} {\bibinfo {title} {Cosmic structure as the quantum interference of a coherent dark wave},\ }\href {https://doi.org/10.1038/nphys2996} {\bibfield  {journal} {\bibinfo  {journal} {Nature Physics}\ }\textbf {\bibinfo {volume} {10}},\ \bibinfo {pages} {496–499} (\bibinfo {year} {2014}{\natexlab{a}})}\BibitemShut {NoStop}%
\bibitem [{\citenamefont {Schive}\ \emph {et~al.}(2014{\natexlab{b}})\citenamefont {Schive}, \citenamefont {Liao}, \citenamefont {Woo}, \citenamefont {Wong}, \citenamefont {Chiueh}, \citenamefont {Broadhurst},\ and\ \citenamefont {Hwang}}]{Schive_2014}%
  \BibitemOpen
  \bibfield  {author} {\bibinfo {author} {\bibfnamefont {H.-Y.}\ \bibnamefont {Schive}}, \bibinfo {author} {\bibfnamefont {M.-H.}\ \bibnamefont {Liao}}, \bibinfo {author} {\bibfnamefont {T.-P.}\ \bibnamefont {Woo}}, \bibinfo {author} {\bibfnamefont {S.-K.}\ \bibnamefont {Wong}}, \bibinfo {author} {\bibfnamefont {T.}~\bibnamefont {Chiueh}}, \bibinfo {author} {\bibfnamefont {T.}~\bibnamefont {Broadhurst}},\ and\ \bibinfo {author} {\bibfnamefont {W.-Y.~P.}\ \bibnamefont {Hwang}},\ }\bibfield  {title} {\bibinfo {title} {{Understanding the Core-Halo Relation of Quantum Wave Dark Matter from 3D Simulations}},\ }\bibfield  {journal} {\bibinfo  {journal} {Physical Review Letters}\ }\textbf {\bibinfo {volume} {113}},\ \href {https://doi.org/10.1103/physrevlett.113.261302} {10.1103/physrevlett.113.261302} (\bibinfo {year} {2014}{\natexlab{b}})\BibitemShut {NoStop}%
\bibitem [{\citenamefont {Chavanis}(2011)}]{Chavanis_2011}%
  \BibitemOpen
  \bibfield  {author} {\bibinfo {author} {\bibfnamefont {P.-H.}\ \bibnamefont {Chavanis}},\ }\bibfield  {title} {\bibinfo {title} {{Mass-radius relation of Newtonian self-gravitating Bose-Einstein condensates with short-range interactions. I. Analytical results}},\ }\bibfield  {journal} {\bibinfo  {journal} {Physical Review D}\ }\textbf {\bibinfo {volume} {84}},\ \href {https://doi.org/10.1103/physrevd.84.043531} {10.1103/physrevd.84.043531} (\bibinfo {year} {2011})\BibitemShut {NoStop}%
\bibitem [{\citenamefont {Navarro}\ \emph {et~al.}(1997)\citenamefont {Navarro}, \citenamefont {Frenk},\ and\ \citenamefont {White}}]{Navarro_1997}%
  \BibitemOpen
  \bibfield  {author} {\bibinfo {author} {\bibfnamefont {J.~F.}\ \bibnamefont {Navarro}}, \bibinfo {author} {\bibfnamefont {C.~S.}\ \bibnamefont {Frenk}},\ and\ \bibinfo {author} {\bibfnamefont {S.~D.~M.}\ \bibnamefont {White}},\ }\bibfield  {title} {\bibinfo {title} {{A Universal Density Profile from Hierarchical Clustering}},\ }\href {https://doi.org/10.1086/304888} {\bibfield  {journal} {\bibinfo  {journal} {The Astrophysical Journal}\ }\textbf {\bibinfo {volume} {490}},\ \bibinfo {pages} {493–508} (\bibinfo {year} {1997})}\BibitemShut {NoStop}%
\bibitem [{\citenamefont {Li}\ \emph {et~al.}(2021)\citenamefont {Li}, \citenamefont {Hui},\ and\ \citenamefont {Yavetz}}]{Li_2021}%
  \BibitemOpen
  \bibfield  {author} {\bibinfo {author} {\bibfnamefont {X.}~\bibnamefont {Li}}, \bibinfo {author} {\bibfnamefont {L.}~\bibnamefont {Hui}},\ and\ \bibinfo {author} {\bibfnamefont {T.~D.}\ \bibnamefont {Yavetz}},\ }\bibfield  {title} {\bibinfo {title} {Oscillations and random walk of the soliton core in a fuzzy dark matter halo},\ }\bibfield  {journal} {\bibinfo  {journal} {Physical Review D}\ }\textbf {\bibinfo {volume} {103}},\ \href {https://doi.org/10.1103/physrevd.103.023508} {10.1103/physrevd.103.023508} (\bibinfo {year} {2021})\BibitemShut {NoStop}%
\bibitem [{\citenamefont {Liu}\ \emph {et~al.}(2023)\citenamefont {Liu}, \citenamefont {Proukakis},\ and\ \citenamefont {Rigopoulos}}]{Liu_2023}%
  \BibitemOpen
  \bibfield  {author} {\bibinfo {author} {\bibfnamefont {I.-K.}\ \bibnamefont {Liu}}, \bibinfo {author} {\bibfnamefont {N.~P.}\ \bibnamefont {Proukakis}},\ and\ \bibinfo {author} {\bibfnamefont {G.}~\bibnamefont {Rigopoulos}},\ }\bibfield  {title} {\bibinfo {title} {Coherent and incoherent structures in fuzzy dark matter haloes},\ }\href {https://doi.org/10.1093/mnras/stad591} {\bibfield  {journal} {\bibinfo  {journal} {Monthly Notices of the Royal Astronomical Society}\ }\textbf {\bibinfo {volume} {521}},\ \bibinfo {pages} {3625–3647} (\bibinfo {year} {2023})}\BibitemShut {NoStop}%
\bibitem [{\citenamefont {Veltmaat}\ \emph {et~al.}(2018)\citenamefont {Veltmaat}, \citenamefont {Niemeyer},\ and\ \citenamefont {Schwabe}}]{Veltmaat_2018}%
  \BibitemOpen
  \bibfield  {author} {\bibinfo {author} {\bibfnamefont {J.}~\bibnamefont {Veltmaat}}, \bibinfo {author} {\bibfnamefont {J.~C.}\ \bibnamefont {Niemeyer}},\ and\ \bibinfo {author} {\bibfnamefont {B.}~\bibnamefont {Schwabe}},\ }\bibfield  {title} {\bibinfo {title} {Formation and structure of ultralight bosonic dark matter halos},\ }\bibfield  {journal} {\bibinfo  {journal} {Physical Review D}\ }\textbf {\bibinfo {volume} {98}},\ \href {https://doi.org/10.1103/physrevd.98.043509} {10.1103/physrevd.98.043509} (\bibinfo {year} {2018})\BibitemShut {NoStop}%
\bibitem [{\citenamefont {Schive}\ \emph {et~al.}(2020)\citenamefont {Schive}, \citenamefont {Chiueh},\ and\ \citenamefont {Broadhurst}}]{Schive_2020}%
  \BibitemOpen
  \bibfield  {author} {\bibinfo {author} {\bibfnamefont {H.-Y.}\ \bibnamefont {Schive}}, \bibinfo {author} {\bibfnamefont {T.}~\bibnamefont {Chiueh}},\ and\ \bibinfo {author} {\bibfnamefont {T.}~\bibnamefont {Broadhurst}},\ }\bibfield  {title} {\bibinfo {title} {{Soliton Random Walk and the Cluster-Stripping Problem in Ultralight Dark Matter}},\ }\bibfield  {journal} {\bibinfo  {journal} {Physical Review Letters}\ }\textbf {\bibinfo {volume} {124}},\ \href {https://doi.org/10.1103/physrevlett.124.201301} {10.1103/physrevlett.124.201301} (\bibinfo {year} {2020})\BibitemShut {NoStop}%
\bibitem [{\citenamefont {Bar-Or}\ \emph {et~al.}(2019)\citenamefont {Bar-Or}, \citenamefont {Fouvry},\ and\ \citenamefont {Tremaine}}]{Bar_Or_2019}%
  \BibitemOpen
  \bibfield  {author} {\bibinfo {author} {\bibfnamefont {B.}~\bibnamefont {Bar-Or}}, \bibinfo {author} {\bibfnamefont {J.-B.}\ \bibnamefont {Fouvry}},\ and\ \bibinfo {author} {\bibfnamefont {S.}~\bibnamefont {Tremaine}},\ }\bibfield  {title} {\bibinfo {title} {{Relaxation in a Fuzzy Dark Matter Halo}},\ }\href {https://doi.org/10.3847/1538-4357/aaf28c} {\bibfield  {journal} {\bibinfo  {journal} {The Astrophysical Journal}\ }\textbf {\bibinfo {volume} {871}},\ \bibinfo {pages} {28} (\bibinfo {year} {2019})}\BibitemShut {NoStop}%
\bibitem [{\citenamefont {El-Zant}\ \emph {et~al.}(2019)\citenamefont {El-Zant}, \citenamefont {Freundlich}, \citenamefont {Combes},\ and\ \citenamefont {Halle}}]{El_Zant_2019}%
  \BibitemOpen
  \bibfield  {author} {\bibinfo {author} {\bibfnamefont {A.~A.}\ \bibnamefont {El-Zant}}, \bibinfo {author} {\bibfnamefont {J.}~\bibnamefont {Freundlich}}, \bibinfo {author} {\bibfnamefont {F.}~\bibnamefont {Combes}},\ and\ \bibinfo {author} {\bibfnamefont {A.}~\bibnamefont {Halle}},\ }\bibfield  {title} {\bibinfo {title} {The effect of fluctuating fuzzy axion haloes on stellar dynamics: a stochastic model},\ }\href {https://doi.org/10.1093/mnras/stz3478} {\bibfield  {journal} {\bibinfo  {journal} {Monthly Notices of the Royal Astronomical Society}\ }\textbf {\bibinfo {volume} {492}},\ \bibinfo {pages} {877–894} (\bibinfo {year} {2019})}\BibitemShut {NoStop}%
\bibitem [{\citenamefont {Chavanis}(2020)}]{chavanis2020landau}%
  \BibitemOpen
  \bibfield  {author} {\bibinfo {author} {\bibfnamefont {P.-H.}\ \bibnamefont {Chavanis}},\ }\href@noop {} {\bibinfo {title} {{Landau equation for self-gravitating classical and quantum particles: Application to dark matter}}} (\bibinfo {year} {2020}),\ \Eprint {https://arxiv.org/abs/2012.12858} {arXiv:2012.12858 [astro-ph.GA]} \BibitemShut {NoStop}%
\bibitem [{\citenamefont {Dutta~Chowdhury}\ \emph {et~al.}(2023)\citenamefont {Dutta~Chowdhury}, \citenamefont {van~den Bosch}, \citenamefont {van Dokkum}, \citenamefont {Robles}, \citenamefont {Schive},\ and\ \citenamefont {Chiueh}}]{Dutta_Chowdhury_2023}%
  \BibitemOpen
  \bibfield  {author} {\bibinfo {author} {\bibfnamefont {D.}~\bibnamefont {Dutta~Chowdhury}}, \bibinfo {author} {\bibfnamefont {F.~C.}\ \bibnamefont {van~den Bosch}}, \bibinfo {author} {\bibfnamefont {P.}~\bibnamefont {van Dokkum}}, \bibinfo {author} {\bibfnamefont {V.~H.}\ \bibnamefont {Robles}}, \bibinfo {author} {\bibfnamefont {H.-Y.}\ \bibnamefont {Schive}},\ and\ \bibinfo {author} {\bibfnamefont {T.}~\bibnamefont {Chiueh}},\ }\bibfield  {title} {\bibinfo {title} {{On the Dynamical Heating of Dwarf Galaxies in a Fuzzy Dark Matter Halo}},\ }\href {https://doi.org/10.3847/1538-4357/acc73d} {\bibfield  {journal} {\bibinfo  {journal} {The Astrophysical Journal}\ }\textbf {\bibinfo {volume} {949}},\ \bibinfo {pages} {68} (\bibinfo {year} {2023})}\BibitemShut {NoStop}%
\bibitem [{\citenamefont {Dutta~Chowdhury}\ \emph {et~al.}(2021)\citenamefont {Dutta~Chowdhury}, \citenamefont {van~den Bosch}, \citenamefont {Robles}, \citenamefont {van Dokkum}, \citenamefont {Schive}, \citenamefont {Chiueh},\ and\ \citenamefont {Broadhurst}}]{Dutta_Chowdhury_2021}%
  \BibitemOpen
  \bibfield  {author} {\bibinfo {author} {\bibfnamefont {D.}~\bibnamefont {Dutta~Chowdhury}}, \bibinfo {author} {\bibfnamefont {F.~C.}\ \bibnamefont {van~den Bosch}}, \bibinfo {author} {\bibfnamefont {V.~H.}\ \bibnamefont {Robles}}, \bibinfo {author} {\bibfnamefont {P.}~\bibnamefont {van Dokkum}}, \bibinfo {author} {\bibfnamefont {H.-Y.}\ \bibnamefont {Schive}}, \bibinfo {author} {\bibfnamefont {T.}~\bibnamefont {Chiueh}},\ and\ \bibinfo {author} {\bibfnamefont {T.}~\bibnamefont {Broadhurst}},\ }\bibfield  {title} {\bibinfo {title} {{On the Random Motion of Nuclear Objects in a Fuzzy Dark Matter Halo}},\ }\href {https://doi.org/10.3847/1538-4357/ac043f} {\bibfield  {journal} {\bibinfo  {journal} {The Astrophysical Journal}\ }\textbf {\bibinfo {volume} {916}},\ \bibinfo {pages} {27} (\bibinfo {year} {2021})}\BibitemShut {NoStop}%
\bibitem [{\citenamefont {Marsh}\ and\ \citenamefont {Niemeyer}(2019)}]{Marsh_2019}%
  \BibitemOpen
  \bibfield  {author} {\bibinfo {author} {\bibfnamefont {D.~J.~E.}\ \bibnamefont {Marsh}}\ and\ \bibinfo {author} {\bibfnamefont {J.~C.}\ \bibnamefont {Niemeyer}},\ }\bibfield  {title} {\bibinfo {title} {{Strong Constraints on Fuzzy Dark Matter from Ultrafaint Dwarf Galaxy Eridanus II}},\ }\bibfield  {journal} {\bibinfo  {journal} {Physical Review Letters}\ }\textbf {\bibinfo {volume} {123}},\ \href {https://doi.org/10.1103/physrevlett.123.051103} {10.1103/physrevlett.123.051103} (\bibinfo {year} {2019})\BibitemShut {NoStop}%
\bibitem [{\citenamefont {Church}\ \emph {et~al.}(2019)\citenamefont {Church}, \citenamefont {Mocz},\ and\ \citenamefont {Ostriker}}]{Church_2019}%
  \BibitemOpen
  \bibfield  {author} {\bibinfo {author} {\bibfnamefont {B.~V.}\ \bibnamefont {Church}}, \bibinfo {author} {\bibfnamefont {P.}~\bibnamefont {Mocz}},\ and\ \bibinfo {author} {\bibfnamefont {J.~P.}\ \bibnamefont {Ostriker}},\ }\bibfield  {title} {\bibinfo {title} {Heating of milky way disc stars by dark matter fluctuations in cold dark matter and fuzzy dark matter paradigms},\ }\href {https://doi.org/10.1093/mnras/stz534} {\bibfield  {journal} {\bibinfo  {journal} {Monthly Notices of the Royal Astronomical Society}\ }\textbf {\bibinfo {volume} {485}},\ \bibinfo {pages} {2861–2876} (\bibinfo {year} {2019})}\BibitemShut {NoStop}%
\bibitem [{\citenamefont {{Dalal}}\ and\ \citenamefont {{Kravtsov}}(2022)}]{2022PhRvD.106f3517D}%
  \BibitemOpen
  \bibfield  {author} {\bibinfo {author} {\bibfnamefont {N.}~\bibnamefont {{Dalal}}}\ and\ \bibinfo {author} {\bibfnamefont {A.}~\bibnamefont {{Kravtsov}}},\ }\bibfield  {title} {\bibinfo {title} {{Excluding fuzzy dark matter with sizes and stellar kinematics of ultrafaint dwarf galaxies}},\ }\href {https://doi.org/10.1103/PhysRevD.106.063517} {\bibfield  {journal} {\bibinfo  {journal} {\prd}\ }\textbf {\bibinfo {volume} {106}},\ \bibinfo {eid} {063517} (\bibinfo {year} {2022})}\BibitemShut {NoStop}%
\bibitem [{\citenamefont {Chiang}\ \emph {et~al.}(2021)\citenamefont {Chiang}, \citenamefont {Schive},\ and\ \citenamefont {Chiueh}}]{Chiang_2021}%
  \BibitemOpen
  \bibfield  {author} {\bibinfo {author} {\bibfnamefont {B.~T.}\ \bibnamefont {Chiang}}, \bibinfo {author} {\bibfnamefont {H.-Y.}\ \bibnamefont {Schive}},\ and\ \bibinfo {author} {\bibfnamefont {T.}~\bibnamefont {Chiueh}},\ }\bibfield  {title} {\bibinfo {title} {{Soliton oscillations and revised constraints from Eridanus II of fuzzy dark matter}},\ }\bibfield  {journal} {\bibinfo  {journal} {Physical Review D}\ }\textbf {\bibinfo {volume} {103}},\ \href {https://doi.org/10.1103/physrevd.103.103019} {10.1103/physrevd.103.103019} (\bibinfo {year} {2021})\BibitemShut {NoStop}%
\bibitem [{\citenamefont {{Plummer}}(1911)}]{1911MNRAS..71..460P}%
  \BibitemOpen
  \bibfield  {author} {\bibinfo {author} {\bibfnamefont {H.~C.}\ \bibnamefont {{Plummer}}},\ }\bibfield  {title} {\bibinfo {title} {{On the problem of distribution in globular star clusters}},\ }\href {https://doi.org/10.1093/mnras/71.5.460} {\bibfield  {journal} {\bibinfo  {journal} {Monthly Notices of the Royal Astronomical Society}\ }\textbf {\bibinfo {volume} {71}},\ \bibinfo {pages} {460} (\bibinfo {year} {1911})}\BibitemShut {NoStop}%
\bibitem [{\citenamefont {{Spekkens}}\ and\ \citenamefont {{Karunakaran}}(2018)}]{2018ApJ...855...28S}%
  \BibitemOpen
  \bibfield  {author} {\bibinfo {author} {\bibfnamefont {K.}~\bibnamefont {{Spekkens}}}\ and\ \bibinfo {author} {\bibfnamefont {A.}~\bibnamefont {{Karunakaran}}},\ }\bibfield  {title} {\bibinfo {title} {{Atomic Gas in Blue Ultra Diffuse Galaxies around Hickson Compact Groups}},\ }\href {https://doi.org/10.3847/1538-4357/aa94be} {\bibfield  {journal} {\bibinfo  {journal} {\apj}\ }\textbf {\bibinfo {volume} {855}},\ \bibinfo {eid} {28} (\bibinfo {year} {2018})},\ \Eprint {https://arxiv.org/abs/1710.06557} {arXiv:1710.06557 [astro-ph.GA]} \BibitemShut {NoStop}%
\bibitem [{\citenamefont {Di~Cintio}\ \emph {et~al.}(2016)\citenamefont {Di~Cintio}, \citenamefont {Brook}, \citenamefont {Dutton}, \citenamefont {Macciò}, \citenamefont {Obreja},\ and\ \citenamefont {Dekel}}]{di2017nihao}%
  \BibitemOpen
  \bibfield  {author} {\bibinfo {author} {\bibfnamefont {A.}~\bibnamefont {Di~Cintio}}, \bibinfo {author} {\bibfnamefont {C.~B.}\ \bibnamefont {Brook}}, \bibinfo {author} {\bibfnamefont {A.~A.}\ \bibnamefont {Dutton}}, \bibinfo {author} {\bibfnamefont {A.~V.}\ \bibnamefont {Macciò}}, \bibinfo {author} {\bibfnamefont {A.}~\bibnamefont {Obreja}},\ and\ \bibinfo {author} {\bibfnamefont {A.}~\bibnamefont {Dekel}},\ }\bibfield  {title} {\bibinfo {title} {{{NIHAO – XI}. Formation of ultra-diffuse galaxies by outflows}},\ }\href {https://doi.org/10.1093/mnrasl/slw210} {\bibfield  {journal} {\bibinfo  {journal} {Monthly Notices of the Royal Astronomical Society: Letters}\ }\textbf {\bibinfo {volume} {466}},\ \bibinfo {pages} {L1–L6} (\bibinfo {year} {2016})}\BibitemShut {NoStop}%
\bibitem [{\citenamefont {Chan}\ \emph {et~al.}(2018)\citenamefont {Chan}, \citenamefont {Kereš}, \citenamefont {Wetzel}, \citenamefont {Hopkins}, \citenamefont {Faucher-Giguère}, \citenamefont {El-Badry}, \citenamefont {Garrison-Kimmel},\ and\ \citenamefont {Boylan-Kolchin}}]{chan2018origin}%
  \BibitemOpen
  \bibfield  {author} {\bibinfo {author} {\bibfnamefont {T.~K.}\ \bibnamefont {Chan}}, \bibinfo {author} {\bibfnamefont {D.}~\bibnamefont {Kereš}}, \bibinfo {author} {\bibfnamefont {A.}~\bibnamefont {Wetzel}}, \bibinfo {author} {\bibfnamefont {P.~F.}\ \bibnamefont {Hopkins}}, \bibinfo {author} {\bibfnamefont {C.-A.}\ \bibnamefont {Faucher-Giguère}}, \bibinfo {author} {\bibfnamefont {K.}~\bibnamefont {El-Badry}}, \bibinfo {author} {\bibfnamefont {S.}~\bibnamefont {Garrison-Kimmel}},\ and\ \bibinfo {author} {\bibfnamefont {M.}~\bibnamefont {Boylan-Kolchin}},\ }\bibfield  {title} {\bibinfo {title} {The origin of ultra diffuse galaxies: stellar feedback and quenching},\ }\href {https://doi.org/10.1093/mnras/sty1153} {\bibfield  {journal} {\bibinfo  {journal} {Monthly Notices of the Royal Astronomical Society}\ }\textbf {\bibinfo {volume} {478}},\ \bibinfo {pages} {906–925} (\bibinfo {year} {2018})}\BibitemShut {NoStop}%
\bibitem [{\citenamefont {Carleton}\ \emph {et~al.}(2019)\citenamefont {Carleton}, \citenamefont {Errani}, \citenamefont {Cooper}, \citenamefont {Kaplinghat}, \citenamefont {Peñarrubia},\ and\ \citenamefont {Guo}}]{carleton2019formation}%
  \BibitemOpen
  \bibfield  {author} {\bibinfo {author} {\bibfnamefont {T.}~\bibnamefont {Carleton}}, \bibinfo {author} {\bibfnamefont {R.}~\bibnamefont {Errani}}, \bibinfo {author} {\bibfnamefont {M.}~\bibnamefont {Cooper}}, \bibinfo {author} {\bibfnamefont {M.}~\bibnamefont {Kaplinghat}}, \bibinfo {author} {\bibfnamefont {J.}~\bibnamefont {Peñarrubia}},\ and\ \bibinfo {author} {\bibfnamefont {Y.}~\bibnamefont {Guo}},\ }\bibfield  {title} {\bibinfo {title} {The formation of ultra-diffuse galaxies in cored dark matter haloes through tidal stripping and heating},\ }\href {https://doi.org/10.1093/mnras/stz383} {\bibfield  {journal} {\bibinfo  {journal} {Monthly Notices of the Royal Astronomical Society}\ }\textbf {\bibinfo {volume} {485}},\ \bibinfo {pages} {382–395} (\bibinfo {year} {2019})}\BibitemShut {NoStop}%
\bibitem [{\citenamefont {Jiang}\ \emph {et~al.}(2019)\citenamefont {Jiang}, \citenamefont {Dekel}, \citenamefont {Freundlich}, \citenamefont {Romanowsky}, \citenamefont {Dutton}, \citenamefont {Macciò},\ and\ \citenamefont {Di~Cintio}}]{jiang2019formation}%
  \BibitemOpen
  \bibfield  {author} {\bibinfo {author} {\bibfnamefont {F.}~\bibnamefont {Jiang}}, \bibinfo {author} {\bibfnamefont {A.}~\bibnamefont {Dekel}}, \bibinfo {author} {\bibfnamefont {J.}~\bibnamefont {Freundlich}}, \bibinfo {author} {\bibfnamefont {A.~J.}\ \bibnamefont {Romanowsky}}, \bibinfo {author} {\bibfnamefont {A.~A.}\ \bibnamefont {Dutton}}, \bibinfo {author} {\bibfnamefont {A.~V.}\ \bibnamefont {Macciò}},\ and\ \bibinfo {author} {\bibfnamefont {A.}~\bibnamefont {Di~Cintio}},\ }\bibfield  {title} {\bibinfo {title} {Formation of ultra-diffuse galaxies in the field and in galaxy groups},\ }\href {https://doi.org/10.1093/mnras/stz1499} {\bibfield  {journal} {\bibinfo  {journal} {Monthly Notices of the Royal Astronomical Society}\ }\textbf {\bibinfo {volume} {487}},\ \bibinfo {pages} {5272–5290} (\bibinfo {year} {2019})}\BibitemShut {NoStop}%
\bibitem [{\citenamefont {Sales}\ \emph {et~al.}(2020)\citenamefont {Sales}, \citenamefont {Navarro}, \citenamefont {Peñafiel}, \citenamefont {Peng}, \citenamefont {Lim},\ and\ \citenamefont {Hernquist}}]{sales2020formation}%
  \BibitemOpen
  \bibfield  {author} {\bibinfo {author} {\bibfnamefont {L.~V.}\ \bibnamefont {Sales}}, \bibinfo {author} {\bibfnamefont {J.~F.}\ \bibnamefont {Navarro}}, \bibinfo {author} {\bibfnamefont {L.}~\bibnamefont {Peñafiel}}, \bibinfo {author} {\bibfnamefont {E.~W.}\ \bibnamefont {Peng}}, \bibinfo {author} {\bibfnamefont {S.}~\bibnamefont {Lim}},\ and\ \bibinfo {author} {\bibfnamefont {L.}~\bibnamefont {Hernquist}},\ }\bibfield  {title} {\bibinfo {title} {The formation of ultradiffuse galaxies in clusters},\ }\href {https://doi.org/10.1093/mnras/staa854} {\bibfield  {journal} {\bibinfo  {journal} {Monthly Notices of the Royal Astronomical Society}\ }\textbf {\bibinfo {volume} {494}},\ \bibinfo {pages} {1848–1858} (\bibinfo {year} {2020})}\BibitemShut {NoStop}%
\bibitem [{\citenamefont {Amorisco}\ and\ \citenamefont {Loeb}(2016)}]{amorisco2016ultradiffuse}%
  \BibitemOpen
  \bibfield  {author} {\bibinfo {author} {\bibfnamefont {N.~C.}\ \bibnamefont {Amorisco}}\ and\ \bibinfo {author} {\bibfnamefont {A.}~\bibnamefont {Loeb}},\ }\bibfield  {title} {\bibinfo {title} {Ultradiffuse galaxies: the high-spin tail of the abundant dwarf galaxy population},\ }\href {https://doi.org/10.1093/mnrasl/slw055} {\bibfield  {journal} {\bibinfo  {journal} {Monthly Notices of the Royal Astronomical Society: Letters}\ }\textbf {\bibinfo {volume} {459}},\ \bibinfo {pages} {L51–L55} (\bibinfo {year} {2016})}\BibitemShut {NoStop}%
\bibitem [{\citenamefont {Guzman}\ and\ \citenamefont {Urena‐Lopez}(2006)}]{Guzman_2006}%
  \BibitemOpen
  \bibfield  {author} {\bibinfo {author} {\bibfnamefont {F.~S.}\ \bibnamefont {Guzman}}\ and\ \bibinfo {author} {\bibfnamefont {L.~A.}\ \bibnamefont {Urena‐Lopez}},\ }\bibfield  {title} {\bibinfo {title} {{Gravitational Cooling of Self‐gravitating Bose Condensates}},\ }\href {https://doi.org/10.1086/504508} {\bibfield  {journal} {\bibinfo  {journal} {The Astrophysical Journal}\ }\textbf {\bibinfo {volume} {645}},\ \bibinfo {pages} {814–819} (\bibinfo {year} {2006})}\BibitemShut {NoStop}%
\bibitem [{\citenamefont {Mocz}\ \emph {et~al.}(2017)\citenamefont {Mocz}, \citenamefont {Vogelsberger}, \citenamefont {Robles}, \citenamefont {Zavala}, \citenamefont {Boylan-Kolchin}, \citenamefont {Fialkov},\ and\ \citenamefont {Hernquist}}]{Mocz_2017}%
  \BibitemOpen
  \bibfield  {author} {\bibinfo {author} {\bibfnamefont {P.}~\bibnamefont {Mocz}}, \bibinfo {author} {\bibfnamefont {M.}~\bibnamefont {Vogelsberger}}, \bibinfo {author} {\bibfnamefont {V.~H.}\ \bibnamefont {Robles}}, \bibinfo {author} {\bibfnamefont {J.}~\bibnamefont {Zavala}}, \bibinfo {author} {\bibfnamefont {M.}~\bibnamefont {Boylan-Kolchin}}, \bibinfo {author} {\bibfnamefont {A.}~\bibnamefont {Fialkov}},\ and\ \bibinfo {author} {\bibfnamefont {L.}~\bibnamefont {Hernquist}},\ }\bibfield  {title} {\bibinfo {title} {{Galaxy formation with BECDM – I. Turbulence and relaxation of idealized haloes}},\ }\href {https://doi.org/10.1093/mnras/stx1887} {\bibfield  {journal} {\bibinfo  {journal} {Monthly Notices of the Royal Astronomical Society}\ }\textbf {\bibinfo {volume} {471}},\ \bibinfo {pages} {4559–4570} (\bibinfo {year} {2017})}\BibitemShut {NoStop}%
\bibitem [{\citenamefont {Hui}\ \emph {et~al.}(2021)\citenamefont {Hui}, \citenamefont {Joyce}, \citenamefont {Landry},\ and\ \citenamefont {Li}}]{Hui_2021}%
  \BibitemOpen
  \bibfield  {author} {\bibinfo {author} {\bibfnamefont {L.}~\bibnamefont {Hui}}, \bibinfo {author} {\bibfnamefont {A.}~\bibnamefont {Joyce}}, \bibinfo {author} {\bibfnamefont {M.~J.}\ \bibnamefont {Landry}},\ and\ \bibinfo {author} {\bibfnamefont {X.}~\bibnamefont {Li}},\ }\bibfield  {title} {\bibinfo {title} {Vortices and waves in light dark matter},\ }\href {https://doi.org/10.1088/1475-7516/2021/01/011} {\bibfield  {journal} {\bibinfo  {journal} {Journal of Cosmology and Astroparticle Physics}\ }\textbf {\bibinfo {volume} {2021}}\bibinfo  {number} { (01)},\ \bibinfo {pages} {011–011}}\BibitemShut {NoStop}%
\bibitem [{\citenamefont {Chiang}\ \emph {et~al.}(2022)\citenamefont {Chiang}, \citenamefont {Ostriker},\ and\ \citenamefont {Schive}}]{Chiang_2022}%
  \BibitemOpen
\bibfield  {number} {  }\bibfield  {author} {\bibinfo {author} {\bibfnamefont {B.~T.}\ \bibnamefont {Chiang}}, \bibinfo {author} {\bibfnamefont {J.~P.}\ \bibnamefont {Ostriker}},\ and\ \bibinfo {author} {\bibfnamefont {H.-Y.}\ \bibnamefont {Schive}},\ }\bibfield  {title} {\bibinfo {title} {{Can ultralight dark matter explain the age–velocity dispersion relation of the Milky Way disc: A revised and improved treatment}},\ }\href {https://doi.org/10.1093/mnras/stac3358} {\bibfield  {journal} {\bibinfo  {journal} {Monthly Notices of the Royal Astronomical Society}\ }\textbf {\bibinfo {volume} {518}},\ \bibinfo {pages} {4045–4063} (\bibinfo {year} {2022})}\BibitemShut {NoStop}%
\bibitem [{\citenamefont {Geringer-Sameth}\ \emph {et~al.}(2015)\citenamefont {Geringer-Sameth}, \citenamefont {Koushiappas},\ and\ \citenamefont {Walker}}]{Geringer_Sameth_2015}%
  \BibitemOpen
  \bibfield  {author} {\bibinfo {author} {\bibfnamefont {A.}~\bibnamefont {Geringer-Sameth}}, \bibinfo {author} {\bibfnamefont {S.~M.}\ \bibnamefont {Koushiappas}},\ and\ \bibinfo {author} {\bibfnamefont {M.}~\bibnamefont {Walker}},\ }\bibfield  {title} {\bibinfo {title} {Dwarf galaxy annihilation and decay emission profiles for dark matter experiments},\ }\href {https://doi.org/10.1088/0004-637x/801/2/74} {\bibfield  {journal} {\bibinfo  {journal} {The Astrophysical Journal}\ }\textbf {\bibinfo {volume} {801}},\ \bibinfo {pages} {74} (\bibinfo {year} {2015})}\BibitemShut {NoStop}%
\bibitem [{\citenamefont {Koushiappas}\ and\ \citenamefont {Loeb}(2017)}]{Koushiappas_2017}%
  \BibitemOpen
  \bibfield  {author} {\bibinfo {author} {\bibfnamefont {S.~M.}\ \bibnamefont {Koushiappas}}\ and\ \bibinfo {author} {\bibfnamefont {A.}~\bibnamefont {Loeb}},\ }\bibfield  {title} {\bibinfo {title} {{Dynamics of Dwarf Galaxies Disfavor Stellar-Mass Black Holes as Dark Matter}},\ }\bibfield  {journal} {\bibinfo  {journal} {Physical Review Letters}\ }\textbf {\bibinfo {volume} {119}},\ \href {https://doi.org/10.1103/physrevlett.119.041102} {10.1103/physrevlett.119.041102} (\bibinfo {year} {2017})\BibitemShut {NoStop}%
\bibitem [{\citenamefont {Wolf}\ \emph {et~al.}(2010)\citenamefont {Wolf}, \citenamefont {Martinez}, \citenamefont {Bullock}, \citenamefont {Kaplinghat}, \citenamefont {Geha}, \citenamefont {Muñoz}, \citenamefont {Simon},\ and\ \citenamefont {Avedo}}]{Wolf_2010}%
  \BibitemOpen
  \bibfield  {author} {\bibinfo {author} {\bibfnamefont {J.}~\bibnamefont {Wolf}}, \bibinfo {author} {\bibfnamefont {G.~D.}\ \bibnamefont {Martinez}}, \bibinfo {author} {\bibfnamefont {J.~S.}\ \bibnamefont {Bullock}}, \bibinfo {author} {\bibfnamefont {M.}~\bibnamefont {Kaplinghat}}, \bibinfo {author} {\bibfnamefont {M.}~\bibnamefont {Geha}}, \bibinfo {author} {\bibfnamefont {R.~R.}\ \bibnamefont {Muñoz}}, \bibinfo {author} {\bibfnamefont {J.~D.}\ \bibnamefont {Simon}},\ and\ \bibinfo {author} {\bibfnamefont {F.~F.}\ \bibnamefont {Avedo}},\ }\bibfield  {title} {\bibinfo {title} {{Accurate masses for dispersion-supported galaxies: Accurate masses for spheroidal galaxies}},\ }\href {https://doi.org/10.1111/j.1365-2966.2010.16753.x} {\bibfield  {journal} {\bibinfo  {journal} {Monthly Notices of the Royal Astronomical Society}\ ,\ \bibinfo {pages} {no}} (\bibinfo {year} {2010})}\BibitemShut {NoStop}%
\bibitem [{\citenamefont {Simon}(2019)}]{Simon_2019}%
  \BibitemOpen
  \bibfield  {author} {\bibinfo {author} {\bibfnamefont {J.~D.}\ \bibnamefont {Simon}},\ }\bibfield  {title} {\bibinfo {title} {{The Faintest Dwarf Galaxies}},\ }\href {https://doi.org/10.1146/annurev-astro-091918-104453} {\bibfield  {journal} {\bibinfo  {journal} {Annual Review of Astronomy and Astrophysics}\ }\textbf {\bibinfo {volume} {57}},\ \bibinfo {pages} {375–415} (\bibinfo {year} {2019})}\BibitemShut {NoStop}%
\bibitem [{\citenamefont {Lora}\ \emph {et~al.}(2012)\citenamefont {Lora}, \citenamefont {Magaña}, \citenamefont {Bernal}, \citenamefont {Sánchez-Salcedo},\ and\ \citenamefont {Grebel}}]{Lora_2012}%
  \BibitemOpen
  \bibfield  {author} {\bibinfo {author} {\bibfnamefont {V.}~\bibnamefont {Lora}}, \bibinfo {author} {\bibfnamefont {J.}~\bibnamefont {Magaña}}, \bibinfo {author} {\bibfnamefont {A.}~\bibnamefont {Bernal}}, \bibinfo {author} {\bibfnamefont {F.}~\bibnamefont {Sánchez-Salcedo}},\ and\ \bibinfo {author} {\bibfnamefont {E.}~\bibnamefont {Grebel}},\ }\bibfield  {title} {\bibinfo {title} {On the mass of ultra-light bosonic dark matter from galactic dynamics},\ }\href {https://doi.org/10.1088/1475-7516/2012/02/011} {\bibfield  {journal} {\bibinfo  {journal} {Journal of Cosmology and Astroparticle Physics}\ }\textbf {\bibinfo {volume} {2012}}\bibinfo  {number} { (02)},\ \bibinfo {pages} {011–011}}\BibitemShut {NoStop}%
\bibitem [{\citenamefont {Hložek}\ \emph {et~al.}(2018)\citenamefont {Hložek}, \citenamefont {Marsh},\ and\ \citenamefont {Grin}}]{Hlo_ek_2018}%
  \BibitemOpen
\bibfield  {number} {  }\bibfield  {author} {\bibinfo {author} {\bibfnamefont {R.}~\bibnamefont {Hložek}}, \bibinfo {author} {\bibfnamefont {D.~J.~E.}\ \bibnamefont {Marsh}},\ and\ \bibinfo {author} {\bibfnamefont {D.}~\bibnamefont {Grin}},\ }\bibfield  {title} {\bibinfo {title} {Using the full power of the cosmic microwave background to probe axion dark matter},\ }\href {https://doi.org/10.1093/mnras/sty271} {\bibfield  {journal} {\bibinfo  {journal} {Monthly Notices of the Royal Astronomical Society}\ }\textbf {\bibinfo {volume} {476}},\ \bibinfo {pages} {3063–3085} (\bibinfo {year} {2018})}\BibitemShut {NoStop}%
\bibitem [{\citenamefont {González-Morales}\ \emph {et~al.}(2017)\citenamefont {González-Morales}, \citenamefont {Marsh}, \citenamefont {Peñarrubia},\ and\ \citenamefont {Ureña-López}}]{Gonz_lez_Morales_2017}%
  \BibitemOpen
  \bibfield  {author} {\bibinfo {author} {\bibfnamefont {A.~X.}\ \bibnamefont {González-Morales}}, \bibinfo {author} {\bibfnamefont {D.~J.~E.}\ \bibnamefont {Marsh}}, \bibinfo {author} {\bibfnamefont {J.}~\bibnamefont {Peñarrubia}},\ and\ \bibinfo {author} {\bibfnamefont {L.~A.}\ \bibnamefont {Ureña-López}},\ }\bibfield  {title} {\bibinfo {title} {Unbiased constraints on ultralight axion mass from dwarf spheroidal galaxies},\ }\href {https://doi.org/10.1093/mnras/stx1941} {\bibfield  {journal} {\bibinfo  {journal} {Monthly Notices of the Royal Astronomical Society}\ }\textbf {\bibinfo {volume} {472}},\ \bibinfo {pages} {1346–1360} (\bibinfo {year} {2017})}\BibitemShut {NoStop}%
\bibitem [{\citenamefont {Bañares-Hernández}\ \emph {et~al.}(2023)\citenamefont {Bañares-Hernández}, \citenamefont {Castillo}, \citenamefont {Martin~Camalich},\ and\ \citenamefont {Iorio}}]{Ba_ares_Hern_ndez_2023}%
  \BibitemOpen
  \bibfield  {author} {\bibinfo {author} {\bibfnamefont {A.}~\bibnamefont {Bañares-Hernández}}, \bibinfo {author} {\bibfnamefont {A.}~\bibnamefont {Castillo}}, \bibinfo {author} {\bibfnamefont {J.}~\bibnamefont {Martin~Camalich}},\ and\ \bibinfo {author} {\bibfnamefont {G.}~\bibnamefont {Iorio}},\ }\bibfield  {title} {\bibinfo {title} {Confronting fuzzy dark matter with the rotation curves of nearby dwarf irregular galaxies},\ }\href {https://doi.org/10.1051/0004-6361/202346686} {\bibfield  {journal} {\bibinfo  {journal} {Astronomy $\&$amp; Astrophysics}\ }\textbf {\bibinfo {volume} {676}},\ \bibinfo {pages} {A63} (\bibinfo {year} {2023})}\BibitemShut {NoStop}%
\bibitem [{\citenamefont {Bernal}\ \emph {et~al.}(2017)\citenamefont {Bernal}, \citenamefont {Fernández-Hernández}, \citenamefont {Matos},\ and\ \citenamefont {Rodríguez-Meza}}]{Bernal_2017}%
  \BibitemOpen
  \bibfield  {author} {\bibinfo {author} {\bibfnamefont {T.}~\bibnamefont {Bernal}}, \bibinfo {author} {\bibfnamefont {L.~M.}\ \bibnamefont {Fernández-Hernández}}, \bibinfo {author} {\bibfnamefont {T.}~\bibnamefont {Matos}},\ and\ \bibinfo {author} {\bibfnamefont {M.~A.}\ \bibnamefont {Rodríguez-Meza}},\ }\bibfield  {title} {\bibinfo {title} {{Rotation curves of high-resolution LSB and SPARC galaxies with fuzzy and multistate (ultralight boson) scalar field dark matter}},\ }\href {https://doi.org/10.1093/mnras/stx3208} {\bibfield  {journal} {\bibinfo  {journal} {Monthly Notices of the Royal Astronomical Society}\ }\textbf {\bibinfo {volume} {475}},\ \bibinfo {pages} {1447–1468} (\bibinfo {year} {2017})}\BibitemShut {NoStop}%
\bibitem [{\citenamefont {Paredes}\ and\ \citenamefont {Michinel}(2016)}]{Paredes_2016}%
  \BibitemOpen
  \bibfield  {author} {\bibinfo {author} {\bibfnamefont {A.}~\bibnamefont {Paredes}}\ and\ \bibinfo {author} {\bibfnamefont {H.}~\bibnamefont {Michinel}},\ }\bibfield  {title} {\bibinfo {title} {Interference of dark matter solitons and galactic offsets},\ }\href {https://doi.org/10.1016/j.dark.2016.02.003} {\bibfield  {journal} {\bibinfo  {journal} {Physics of the Dark Universe}\ }\textbf {\bibinfo {volume} {12}},\ \bibinfo {pages} {50–55} (\bibinfo {year} {2016})}\BibitemShut {NoStop}%
\bibitem [{\citenamefont {Armengaud}\ \emph {et~al.}(2017)\citenamefont {Armengaud}, \citenamefont {Palanque-Delabrouille}, \citenamefont {Yèche}, \citenamefont {Marsh},\ and\ \citenamefont {Baur}}]{Armengaud_2017}%
  \BibitemOpen
  \bibfield  {author} {\bibinfo {author} {\bibfnamefont {E.}~\bibnamefont {Armengaud}}, \bibinfo {author} {\bibfnamefont {N.}~\bibnamefont {Palanque-Delabrouille}}, \bibinfo {author} {\bibfnamefont {C.}~\bibnamefont {Yèche}}, \bibinfo {author} {\bibfnamefont {D.~J.~E.}\ \bibnamefont {Marsh}},\ and\ \bibinfo {author} {\bibfnamefont {J.}~\bibnamefont {Baur}},\ }\bibfield  {title} {\bibinfo {title} {{Constraining the mass of light bosonic dark matter using SDSS Lyman-$\alpha$ forest}},\ }\href {https://doi.org/10.1093/mnras/stx1870} {\bibfield  {journal} {\bibinfo  {journal} {Monthly Notices of the Royal Astronomical Society}\ }\textbf {\bibinfo {volume} {471}},\ \bibinfo {pages} {4606–4614} (\bibinfo {year} {2017})}\BibitemShut {NoStop}%
\bibitem [{\citenamefont {Iršič}\ \emph {et~al.}(2017)\citenamefont {Iršič}, \citenamefont {Viel}, \citenamefont {Haehnelt}, \citenamefont {Bolton},\ and\ \citenamefont {Becker}}]{Ir_i__2017}%
  \BibitemOpen
  \bibfield  {author} {\bibinfo {author} {\bibfnamefont {V.}~\bibnamefont {Iršič}}, \bibinfo {author} {\bibfnamefont {M.}~\bibnamefont {Viel}}, \bibinfo {author} {\bibfnamefont {M.~G.}\ \bibnamefont {Haehnelt}}, \bibinfo {author} {\bibfnamefont {J.~S.}\ \bibnamefont {Bolton}},\ and\ \bibinfo {author} {\bibfnamefont {G.~D.}\ \bibnamefont {Becker}},\ }\bibfield  {title} {\bibinfo {title} {{First Constraints on Fuzzy Dark Matter from Lyman-$\alpha$ Forest Data and Hydrodynamical Simulations}},\ }\bibfield  {journal} {\bibinfo  {journal} {Physical Review Letters}\ }\textbf {\bibinfo {volume} {119}},\ \href {https://doi.org/10.1103/physrevlett.119.031302} {10.1103/physrevlett.119.031302} (\bibinfo {year} {2017})\BibitemShut {NoStop}%
\bibitem [{\citenamefont {Kobayashi}\ \emph {et~al.}(2017)\citenamefont {Kobayashi}, \citenamefont {Murgia}, \citenamefont {De~Simone}, \citenamefont {Iršič},\ and\ \citenamefont {Viel}}]{Kobayashi_2017}%
  \BibitemOpen
  \bibfield  {author} {\bibinfo {author} {\bibfnamefont {T.}~\bibnamefont {Kobayashi}}, \bibinfo {author} {\bibfnamefont {R.}~\bibnamefont {Murgia}}, \bibinfo {author} {\bibfnamefont {A.}~\bibnamefont {De~Simone}}, \bibinfo {author} {\bibfnamefont {V.}~\bibnamefont {Iršič}},\ and\ \bibinfo {author} {\bibfnamefont {M.}~\bibnamefont {Viel}},\ }\bibfield  {title} {\bibinfo {title} {{Lyman-$\alpha$ constraints on ultralight scalar dark matter: Implications for the early and late universe}},\ }\bibfield  {journal} {\bibinfo  {journal} {Physical Review D}\ }\textbf {\bibinfo {volume} {96}},\ \href {https://doi.org/10.1103/physrevd.96.123514} {10.1103/physrevd.96.123514} (\bibinfo {year} {2017})\BibitemShut {NoStop}%
\bibitem [{\citenamefont {Nori}\ \emph {et~al.}(2018)\citenamefont {Nori}, \citenamefont {Murgia}, \citenamefont {Iršič}, \citenamefont {Baldi},\ and\ \citenamefont {Viel}}]{Nori_2018}%
  \BibitemOpen
  \bibfield  {author} {\bibinfo {author} {\bibfnamefont {M.}~\bibnamefont {Nori}}, \bibinfo {author} {\bibfnamefont {R.}~\bibnamefont {Murgia}}, \bibinfo {author} {\bibfnamefont {V.}~\bibnamefont {Iršič}}, \bibinfo {author} {\bibfnamefont {M.}~\bibnamefont {Baldi}},\ and\ \bibinfo {author} {\bibfnamefont {M.}~\bibnamefont {Viel}},\ }\bibfield  {title} {\bibinfo {title} {{Lyman-$\alpha$ forest and non-linear structure characterization in Fuzzy Dark Matter cosmologies}},\ }\href {https://doi.org/10.1093/mnras/sty2888} {\bibfield  {journal} {\bibinfo  {journal} {Monthly Notices of the Royal Astronomical Society}\ }\textbf {\bibinfo {volume} {482}},\ \bibinfo {pages} {3227–3243} (\bibinfo {year} {2018})}\BibitemShut {NoStop}%
\bibitem [{\citenamefont {Rogers}\ and\ \citenamefont {Peiris}(2021)}]{Rogers_2021}%
  \BibitemOpen
  \bibfield  {author} {\bibinfo {author} {\bibfnamefont {K.~K.}\ \bibnamefont {Rogers}}\ and\ \bibinfo {author} {\bibfnamefont {H.~V.}\ \bibnamefont {Peiris}},\ }\bibfield  {title} {\bibinfo {title} {{Strong Bound on Canonical Ultralight Axion Dark Matter from the Lyman-Alpha Forest}},\ }\bibfield  {journal} {\bibinfo  {journal} {Physical Review Letters}\ }\textbf {\bibinfo {volume} {126}},\ \href {https://doi.org/10.1103/physrevlett.126.071302} {10.1103/physrevlett.126.071302} (\bibinfo {year} {2021})\BibitemShut {NoStop}%
\bibitem [{\citenamefont {Nadler}\ \emph {et~al.}(2019)\citenamefont {Nadler}, \citenamefont {Gluscevic}, \citenamefont {Boddy},\ and\ \citenamefont {Wechsler}}]{Nadler_2019}%
  \BibitemOpen
  \bibfield  {author} {\bibinfo {author} {\bibfnamefont {E.~O.}\ \bibnamefont {Nadler}}, \bibinfo {author} {\bibfnamefont {V.}~\bibnamefont {Gluscevic}}, \bibinfo {author} {\bibfnamefont {K.~K.}\ \bibnamefont {Boddy}},\ and\ \bibinfo {author} {\bibfnamefont {R.~H.}\ \bibnamefont {Wechsler}},\ }\bibfield  {title} {\bibinfo {title} {{Constraints on Dark Matter Microphysics from the Milky Way Satellite Population}},\ }\href {https://doi.org/10.3847/2041-8213/ab1eb2} {\bibfield  {journal} {\bibinfo  {journal} {The Astrophysical Journal Letters}\ }\textbf {\bibinfo {volume} {878}},\ \bibinfo {pages} {L32} (\bibinfo {year} {2019})}\BibitemShut {NoStop}%
\bibitem [{\citenamefont {Nadler}\ \emph {et~al.}(2021)\citenamefont {Nadler} \emph {et~al.}}]{Nadler_2021}%
  \BibitemOpen
  \bibfield  {author} {\bibinfo {author} {\bibfnamefont {E.~O.}\ \bibnamefont {Nadler}} \emph {et~al.},\ }\bibfield  {title} {\bibinfo {title} {{Constraints on Dark Matter Properties from Observations of Milky Way Satellite Galaxies}},\ }\bibfield  {journal} {\bibinfo  {journal} {Physical Review Letters}\ }\textbf {\bibinfo {volume} {126}},\ \href {https://doi.org/10.1103/physrevlett.126.091101} {10.1103/physrevlett.126.091101} (\bibinfo {year} {2021})\BibitemShut {NoStop}%
\bibitem [{\citenamefont {Benito}\ \emph {et~al.}(2020)\citenamefont {Benito}, \citenamefont {Criado}, \citenamefont {Hütsi}, \citenamefont {Raidal},\ and\ \citenamefont {Veermäe}}]{Benito_2020}%
  \BibitemOpen
  \bibfield  {author} {\bibinfo {author} {\bibfnamefont {M.}~\bibnamefont {Benito}}, \bibinfo {author} {\bibfnamefont {J.~C.}\ \bibnamefont {Criado}}, \bibinfo {author} {\bibfnamefont {G.}~\bibnamefont {Hütsi}}, \bibinfo {author} {\bibfnamefont {M.}~\bibnamefont {Raidal}},\ and\ \bibinfo {author} {\bibfnamefont {H.}~\bibnamefont {Veermäe}},\ }\bibfield  {title} {\bibinfo {title} {{Implications of Milky Way substructures for the nature of dark matter}},\ }\bibfield  {journal} {\bibinfo  {journal} {Physical Review D}\ }\textbf {\bibinfo {volume} {101}},\ \href {https://doi.org/10.1103/physrevd.101.103023} {10.1103/physrevd.101.103023} (\bibinfo {year} {2020})\BibitemShut {NoStop}%
\bibitem [{\citenamefont {Schutz}(2020)}]{Schutz_2020}%
  \BibitemOpen
  \bibfield  {author} {\bibinfo {author} {\bibfnamefont {K.}~\bibnamefont {Schutz}},\ }\bibfield  {title} {\bibinfo {title} {Subhalo mass function and ultralight bosonic dark matter},\ }\bibfield  {journal} {\bibinfo  {journal} {Physical Review D}\ }\textbf {\bibinfo {volume} {101}},\ \href {https://doi.org/10.1103/physrevd.101.123026} {10.1103/physrevd.101.123026} (\bibinfo {year} {2020})\BibitemShut {NoStop}%
\bibitem [{\citenamefont {Ferreira}(2021)}]{Ferreira_2021}%
  \BibitemOpen
  \bibfield  {author} {\bibinfo {author} {\bibfnamefont {E.~G.~M.}\ \bibnamefont {Ferreira}},\ }\bibfield  {title} {\bibinfo {title} {Ultra-light dark matter},\ }\bibfield  {journal} {\bibinfo  {journal} {The Astronomy and Astrophysics Review}\ }\textbf {\bibinfo {volume} {29}},\ \href {https://doi.org/10.1007/s00159-021-00135-6} {10.1007/s00159-021-00135-6} (\bibinfo {year} {2021})\BibitemShut {NoStop}%
\bibitem [{\citenamefont {Prole}\ \emph {et~al.}(2019)\citenamefont {Prole}, \citenamefont {van~der Burg}, \citenamefont {Hilker},\ and\ \citenamefont {Davies}}]{Prole_2019}%
  \BibitemOpen
  \bibfield  {author} {\bibinfo {author} {\bibfnamefont {D.~J.}\ \bibnamefont {Prole}}, \bibinfo {author} {\bibfnamefont {R.~F.~J.}\ \bibnamefont {van~der Burg}}, \bibinfo {author} {\bibfnamefont {M.}~\bibnamefont {Hilker}},\ and\ \bibinfo {author} {\bibfnamefont {J.~I.}\ \bibnamefont {Davies}},\ }\bibfield  {title} {\bibinfo {title} {{Observational properties of ultra-diffuse galaxies in low-density environments: field UDGs are predominantly blue and star forming}},\ }\href {https://doi.org/10.1093/mnras/stz1843} {\bibfield  {journal} {\bibinfo  {journal} {Monthly Notices of the Royal Astronomical Society}\ }\textbf {\bibinfo {volume} {488}},\ \bibinfo {pages} {2143–2157} (\bibinfo {year} {2019})}\BibitemShut {NoStop}%
\bibitem [{\citenamefont {Román}\ and\ \citenamefont {Trujillo}(2017)}]{Rom_n_2017}%
  \BibitemOpen
  \bibfield  {author} {\bibinfo {author} {\bibfnamefont {J.}~\bibnamefont {Román}}\ and\ \bibinfo {author} {\bibfnamefont {I.}~\bibnamefont {Trujillo}},\ }\bibfield  {title} {\bibinfo {title} {Ultra-diffuse galaxies outside clusters: clues to their formation and evolution},\ }\href {https://doi.org/10.1093/mnras/stx694} {\bibfield  {journal} {\bibinfo  {journal} {Monthly Notices of the Royal Astronomical Society}\ }\textbf {\bibinfo {volume} {468}},\ \bibinfo {pages} {4039–4047} (\bibinfo {year} {2017})}\BibitemShut {NoStop}%
\bibitem [{\citenamefont {Jones}\ \emph {et~al.}(2022)\citenamefont {Jones}, \citenamefont {Karunakaran}, \citenamefont {Bennet}, \citenamefont {Sand}, \citenamefont {Spekkens}, \citenamefont {Mutlu-Pakdil}, \citenamefont {Crnojević}, \citenamefont {Janowiecki}, \citenamefont {Leisman},\ and\ \citenamefont {Fielder}}]{Jones_2022}%
  \BibitemOpen
  \bibfield  {author} {\bibinfo {author} {\bibfnamefont {M.~G.}\ \bibnamefont {Jones}}, \bibinfo {author} {\bibfnamefont {A.}~\bibnamefont {Karunakaran}}, \bibinfo {author} {\bibfnamefont {P.}~\bibnamefont {Bennet}}, \bibinfo {author} {\bibfnamefont {D.~J.}\ \bibnamefont {Sand}}, \bibinfo {author} {\bibfnamefont {K.}~\bibnamefont {Spekkens}}, \bibinfo {author} {\bibfnamefont {B.}~\bibnamefont {Mutlu-Pakdil}}, \bibinfo {author} {\bibfnamefont {D.}~\bibnamefont {Crnojević}}, \bibinfo {author} {\bibfnamefont {S.}~\bibnamefont {Janowiecki}}, \bibinfo {author} {\bibfnamefont {L.}~\bibnamefont {Leisman}},\ and\ \bibinfo {author} {\bibfnamefont {C.~E.}\ \bibnamefont {Fielder}},\ }\bibfield  {title} {\bibinfo {title} {{Gas-rich, Field Ultra-diffuse Galaxies Host Few Gobular Clusters}},\ }\href {https://doi.org/10.3847/2041-8213/acaaab} {\bibfield  {journal} {\bibinfo  {journal} {The Astrophysical Journal Letters}\ }\textbf {\bibinfo {volume} {942}},\ \bibinfo {pages} {L5} (\bibinfo {year} {2022})}\BibitemShut
  {NoStop}%
\bibitem [{\citenamefont {Benavides}\ \emph {et~al.}(2023)\citenamefont {Benavides}, \citenamefont {Sales}, \citenamefont {Abadi}, \citenamefont {Marinacci}, \citenamefont {Vogelsberger},\ and\ \citenamefont {Hernquist}}]{Benavides_2023}%
  \BibitemOpen
  \bibfield  {author} {\bibinfo {author} {\bibfnamefont {J.~A.}\ \bibnamefont {Benavides}}, \bibinfo {author} {\bibfnamefont {L.~V.}\ \bibnamefont {Sales}}, \bibinfo {author} {\bibfnamefont {M.~G.}\ \bibnamefont {Abadi}}, \bibinfo {author} {\bibfnamefont {F.}~\bibnamefont {Marinacci}}, \bibinfo {author} {\bibfnamefont {M.}~\bibnamefont {Vogelsberger}},\ and\ \bibinfo {author} {\bibfnamefont {L.}~\bibnamefont {Hernquist}},\ }\bibfield  {title} {\bibinfo {title} {Origin and evolution of ultradiffuse galaxies in different environments},\ }\href {https://doi.org/10.1093/mnras/stad1053} {\bibfield  {journal} {\bibinfo  {journal} {Monthly Notices of the Royal Astronomical Society}\ }\textbf {\bibinfo {volume} {522}},\ \bibinfo {pages} {1033–1048} (\bibinfo {year} {2023})}\BibitemShut {NoStop}%
\bibitem [{\citenamefont {Wang}\ \emph {et~al.}(2016)\citenamefont {Wang}, \citenamefont {Koribalski}, \citenamefont {Serra}, \citenamefont {van~der Hulst}, \citenamefont {Roychowdhury}, \citenamefont {Kamphuis},\ and\ \citenamefont {N.~Chengalur}}]{Wang_2016}%
  \BibitemOpen
  \bibfield  {author} {\bibinfo {author} {\bibfnamefont {J.}~\bibnamefont {Wang}}, \bibinfo {author} {\bibfnamefont {B.~S.}\ \bibnamefont {Koribalski}}, \bibinfo {author} {\bibfnamefont {P.}~\bibnamefont {Serra}}, \bibinfo {author} {\bibfnamefont {T.}~\bibnamefont {van~der Hulst}}, \bibinfo {author} {\bibfnamefont {S.}~\bibnamefont {Roychowdhury}}, \bibinfo {author} {\bibfnamefont {P.}~\bibnamefont {Kamphuis}},\ and\ \bibinfo {author} {\bibfnamefont {J.}~\bibnamefont {N.~Chengalur}},\ }\bibfield  {title} {\bibinfo {title} {New lessons from the {HI} size–mass relation of galaxies},\ }\href {https://doi.org/10.1093/mnras/stw1099} {\bibfield  {journal} {\bibinfo  {journal} {Monthly Notices of the Royal Astronomical Society}\ }\textbf {\bibinfo {volume} {460}},\ \bibinfo {pages} {2143–2151} (\bibinfo {year} {2016})}\BibitemShut {NoStop}%
\bibitem [{\citenamefont {Gault}\ \emph {et~al.}(2021)\citenamefont {Gault}, \citenamefont {Leisman}, \citenamefont {Adams}, \citenamefont {Mancera~Piña}, \citenamefont {Reiter}, \citenamefont {Smith}, \citenamefont {Battipaglia}, \citenamefont {Cannon}, \citenamefont {Fraternali}, \citenamefont {Haynes}, \citenamefont {McAllan}, \citenamefont {Pagel}, \citenamefont {Rhode}, \citenamefont {Salzer},\ and\ \citenamefont {Singer}}]{Gault_2021}%
  \BibitemOpen
  \bibfield  {author} {\bibinfo {author} {\bibfnamefont {L.}~\bibnamefont {Gault}}, \bibinfo {author} {\bibfnamefont {L.}~\bibnamefont {Leisman}}, \bibinfo {author} {\bibfnamefont {E.~A.~K.}\ \bibnamefont {Adams}}, \bibinfo {author} {\bibfnamefont {P.~E.}\ \bibnamefont {Mancera~Piña}}, \bibinfo {author} {\bibfnamefont {K.}~\bibnamefont {Reiter}}, \bibinfo {author} {\bibfnamefont {N.}~\bibnamefont {Smith}}, \bibinfo {author} {\bibfnamefont {M.}~\bibnamefont {Battipaglia}}, \bibinfo {author} {\bibfnamefont {J.~M.}\ \bibnamefont {Cannon}}, \bibinfo {author} {\bibfnamefont {F.}~\bibnamefont {Fraternali}}, \bibinfo {author} {\bibfnamefont {M.~P.}\ \bibnamefont {Haynes}}, \bibinfo {author} {\bibfnamefont {E.}~\bibnamefont {McAllan}}, \bibinfo {author} {\bibfnamefont {H.~J.}\ \bibnamefont {Pagel}}, \bibinfo {author} {\bibfnamefont {K.~L.}\ \bibnamefont {Rhode}}, \bibinfo {author} {\bibfnamefont {J.~J.}\ \bibnamefont {Salzer}},\ and\ \bibinfo {author} {\bibfnamefont {Q.}~\bibnamefont {Singer}},\ }\bibfield  {title}
  {\bibinfo {title} {{{VLA} Imaging of {HI}-bearing Ultra-diffuse Galaxies from the {ALFALFA} Survey}},\ }\href {https://doi.org/10.3847/1538-4357/abd79d} {\bibfield  {journal} {\bibinfo  {journal} {The Astrophysical Journal}\ }\textbf {\bibinfo {volume} {909}},\ \bibinfo {pages} {19} (\bibinfo {year} {2021})}\BibitemShut {NoStop}%
\bibitem [{\citenamefont {Pucha}\ \emph {et~al.}(2019)\citenamefont {Pucha}, \citenamefont {Carlin}, \citenamefont {Willman}, \citenamefont {Strader}, \citenamefont {Sand}, \citenamefont {Bechtol}, \citenamefont {Brodie}, \citenamefont {Crnojević}, \citenamefont {Forbes}, \citenamefont {Garling}, \citenamefont {Hargis}, \citenamefont {Peter},\ and\ \citenamefont {Romanowsky}}]{Pucha_2019}%
  \BibitemOpen
  \bibfield  {author} {\bibinfo {author} {\bibfnamefont {R.}~\bibnamefont {Pucha}}, \bibinfo {author} {\bibfnamefont {J.~L.}\ \bibnamefont {Carlin}}, \bibinfo {author} {\bibfnamefont {B.}~\bibnamefont {Willman}}, \bibinfo {author} {\bibfnamefont {J.}~\bibnamefont {Strader}}, \bibinfo {author} {\bibfnamefont {D.~J.}\ \bibnamefont {Sand}}, \bibinfo {author} {\bibfnamefont {K.}~\bibnamefont {Bechtol}}, \bibinfo {author} {\bibfnamefont {J.~P.}\ \bibnamefont {Brodie}}, \bibinfo {author} {\bibfnamefont {D.}~\bibnamefont {Crnojević}}, \bibinfo {author} {\bibfnamefont {D.~A.}\ \bibnamefont {Forbes}}, \bibinfo {author} {\bibfnamefont {C.}~\bibnamefont {Garling}}, \bibinfo {author} {\bibfnamefont {J.}~\bibnamefont {Hargis}}, \bibinfo {author} {\bibfnamefont {A.~H.~G.}\ \bibnamefont {Peter}},\ and\ \bibinfo {author} {\bibfnamefont {A.~J.}\ \bibnamefont {Romanowsky}},\ }\bibfield  {title} {\bibinfo {title} {{Hyper Wide Field Imaging of the Local Group Dwarf Irregular Galaxy IC 1613: An Extended Component of Metal-poor
  Stars}},\ }\href {https://doi.org/10.3847/1538-4357/ab29fb} {\bibfield  {journal} {\bibinfo  {journal} {The Astrophysical Journal}\ }\textbf {\bibinfo {volume} {880}},\ \bibinfo {pages} {104} (\bibinfo {year} {2019})}\BibitemShut {NoStop}%
\bibitem [{\citenamefont {Higgs}\ \emph {et~al.}(2021)\citenamefont {Higgs}, \citenamefont {McConnachie}, \citenamefont {Annau}, \citenamefont {Irwin}, \citenamefont {Battaglia}, \citenamefont {Côté}, \citenamefont {Lewis},\ and\ \citenamefont {Venn}}]{Higgs_2021}%
  \BibitemOpen
  \bibfield  {author} {\bibinfo {author} {\bibfnamefont {C.~R.}\ \bibnamefont {Higgs}}, \bibinfo {author} {\bibfnamefont {A.~W.}\ \bibnamefont {McConnachie}}, \bibinfo {author} {\bibfnamefont {N.}~\bibnamefont {Annau}}, \bibinfo {author} {\bibfnamefont {M.}~\bibnamefont {Irwin}}, \bibinfo {author} {\bibfnamefont {G.}~\bibnamefont {Battaglia}}, \bibinfo {author} {\bibfnamefont {P.}~\bibnamefont {Côté}}, \bibinfo {author} {\bibfnamefont {G.~F.}\ \bibnamefont {Lewis}},\ and\ \bibinfo {author} {\bibfnamefont {K.}~\bibnamefont {Venn}},\ }\bibfield  {title} {\bibinfo {title} {{Solo dwarfs II: the stellar structure of isolated Local Group dwarf galaxies}},\ }\href {https://doi.org/10.1093/mnras/stab002} {\bibfield  {journal} {\bibinfo  {journal} {Monthly Notices of the Royal Astronomical Society}\ }\textbf {\bibinfo {volume} {503}},\ \bibinfo {pages} {176–199} (\bibinfo {year} {2021})}\BibitemShut {NoStop}%
\bibitem [{\citenamefont {Carleton}\ \emph {et~al.}(2024)\citenamefont {Carleton} \emph {et~al.}}]{carleton2024pearls}%
  \BibitemOpen
  \bibfield  {author} {\bibinfo {author} {\bibfnamefont {T.}~\bibnamefont {Carleton}} \emph {et~al.},\ }\href@noop {} {\bibinfo {title} {{PEARLS: A Potentially Isolated Quiescent Dwarf Galaxy with a TRGB Distance of 30 Mpc}}} (\bibinfo {year} {2024}),\ \Eprint {https://arxiv.org/abs/2309.16028} {arXiv:2309.16028 [astro-ph.GA]} \BibitemShut {NoStop}%
\end{thebibliography}%
\begin{appendix}
    \section{Structures in the final stellar density distribution\label{App}}
    \begin{figure}[htbp]
    \includegraphics[width=0.48\textwidth]{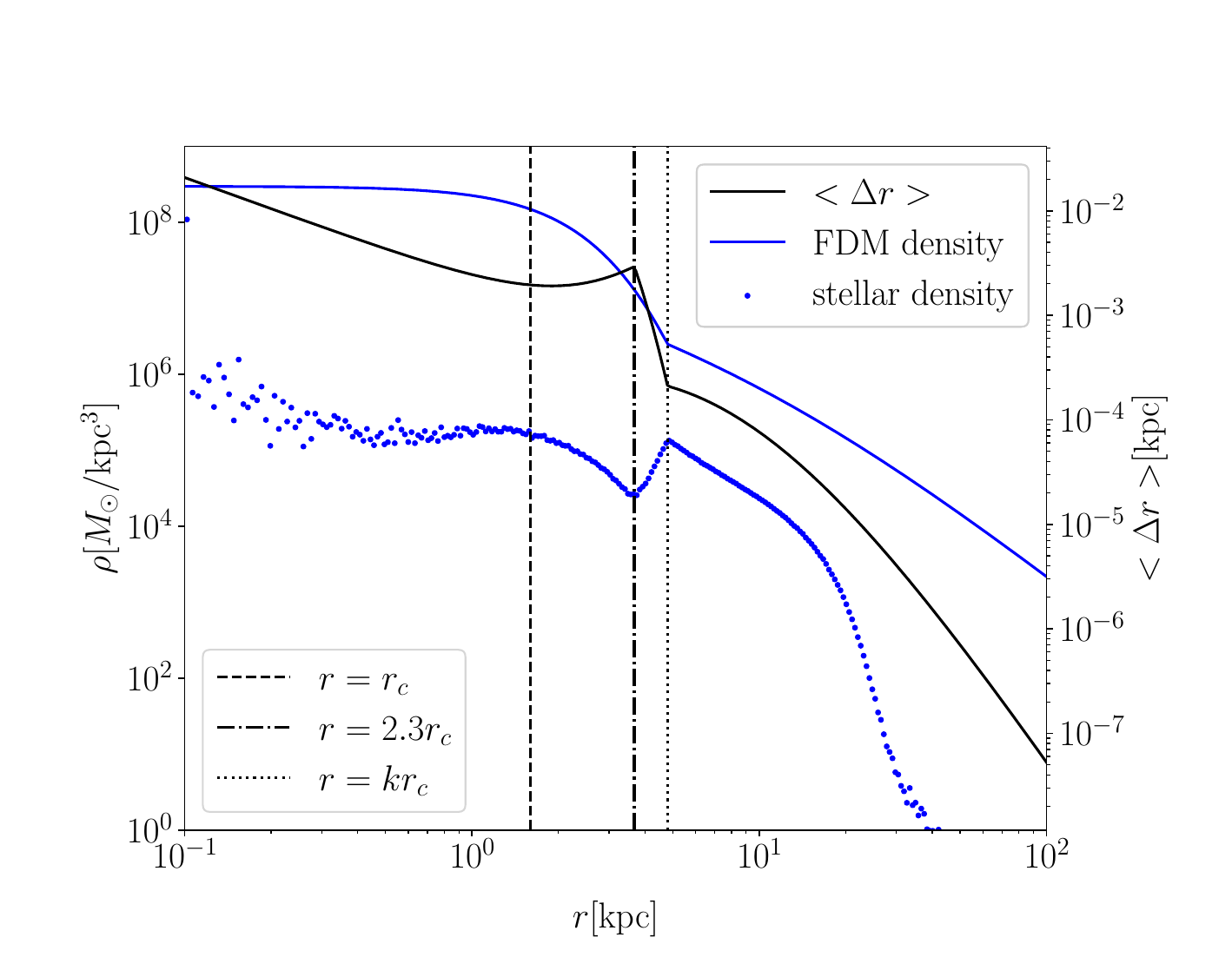}
  \caption{The solid black line illustrates the ensemble-averaged $\Delta r$ within a single timestep for particles located at different distances from the galaxy center. The solid blue line represents the FDM density profile, consistent with those depicted by the blue lines in the upper panels of Fig.\,\ref{figure_2}. The blue dots correspond to the results shown in the lower panels of Fig.\,\ref{figure_2}. The three vertical lines retain the same meaning as in Fig.\,\ref{figure_1}.}
  \label{figure_2.1}
\end{figure}

The final stellar density distribution exhibits a distinct trough followed by a peak, as illustrated in the lower panels of Fig.\,\ref{figure_2}. To further investigate this characteristic, we present the final stellar density distribution corresponding to the parameters $m=1.0\times 10^{-23}$ eV, $k=3$, and $r_s=10$ kpc in Fig.\,\ref{figure_2.1}. Additionally, we display the ensemble-averaged positional changes $\Delta r$  within a single time step, described by Eq.\,\ref{Delta_r} and \ref{averageEM}, for particles located at various distances from the galaxy center. 

In the region $r>2.3r_c$, the profile of $\langle \Delta r\rangle$ has a similar trend to the FDM density distribution, as the energy diffusion coefficient is directly proportional to $\rho^2_\text{FDM}$ in Eq.\,\ref{D_Delta_E}. However, for $r<2.3r_c$, the shape of $\langle \Delta r\rangle$ deviates from the FDM distribution due to the utilization of the effective model proposed in \cite{Dutta_Chowdhury_2021}, where the diffusion coefficient within $2.3r_c$ is taken to be the value at $2.3r_c$. 
The peak of the stellar density can be attributed to particle accumulation resulting from a more pronounced heating effect in the inner region compared to the outer region, where the heating effect diminishes rapidly with increasing $r$. On the other hand, the trough is located at $r=2.3r_c$, where the heating effect derived from the effective model reaches its maximum. Consequently, the presence of this trough can be explained by particles in this region being heated towards the outer regions, while the inner particles near this location do not receive adequate heating effect to redistribute. Therefore, these structures stem from the effective model employed in this analysis and may not be observed in actual simulations.

\end{appendix}
\end{document}